%% file: templateArxiv.tex
\documentclass[twoside]{article}

\usepackage{PRIMEarxiv}

\usepackage[utf8]{inputenc} % allow utf-8 input
\usepackage[T1]{fontenc}    % use 8-bit T1 fonts
\usepackage{hyperref}       % hyperlinks
\usepackage{url}            % simple URL typesetting
\usepackage{xcolor}
\usepackage{multirow}
\usepackage{pifont}
\usepackage{amsthm}
\usepackage{amsmath}
\usepackage{booktabs}       % professional-quality tables
\usepackage{amsfonts}       % blackboard math symbols
\usepackage{nicefrac}       % compact symbols for 1/2, etc.
\usepackage{microtype}      % microtypography
\usepackage{lipsum}
\usepackage{fancyhdr}       % header
\usepackage{graphicx}       % graphics
\graphicspath{{media/}}     % organize your images and other figures under media/ folder

\title{Legal Case Document Similarity: You Need Both Network and Text
\thanks{\textcolor{blue}{This work has been published in Information Processing and Management, Elsevier, vol. 59, issue 6, November 2022.}
}
}

\author{
  Paheli Bhattacharya \\
  Indian Institute of Technology Kharagpur \\
  India \\
  \texttt{paheli@iitkgp.ac.in}
   \And
  Kripabandhu Ghosh \\
  Indian Institute of Science Education and Research, Kolkata \\
  India \\
  \texttt{kripaghosh@iiserkol.ac.in}\\
  \And
  Arindam Pal \\
  Data61, CSIRO and UNSW Sydney \\
  Australia \\
  \texttt{arindamp@gmail.com}
  \And
  Saptarshi Ghosh \\
  Indian Institute of Technology Kharagpur \\
  India \\
  \texttt{saptarshi@cse.iitkgp.ac.in}
 }

\begin{document}
\let\WriteBookmarks\relax
\def\floatpagepagefraction{1}
\def\textpagefraction{.001}

\maketitle

\begin{abstract}
Estimating the similarity between two legal case documents is an important and challenging problem, having various downstream applications such as prior-case retrieval and citation recommendation. 
There are two broad approaches for the task -- citation network-based and text-based. Prior citation network-based approaches consider citations only to prior-cases (also called precedents) (PCNet). This approach misses important signals inherent in Statutes (written laws of a jurisdiction). In this work, we propose Hier-SPCNet that augments PCNet with a heterogeneous network of Statutes. We incorporate domain knowledge for legal document similarity into Hier-SPCNet, thereby obtaining state-of-the-art results for network-based legal document similarity.

Both textual and network similarity provide important signals for legal case similarity; but till now, only trivial attempts have been made to unify the two signals. In this work, we apply several methods for combining textual and network information for estimating legal case similarity. We perform extensive experiments over legal case documents from the Indian judiciary, where the gold standard similarity between document-pairs is judged by law experts from two reputed Law institutes in India. Our experiments establish that our proposed network-based methods significantly improve the correlation with domain experts' opinion when compared to the existing methods for network-based legal document similarity. 
Our best-performing combination method (that combines network-based and text-based similarity) improves the correlation with domain experts' opinion by 11.8\% over the best text-based method and 20.6\% over the best network-based method. 
We also establish that our best-performing method can be used to recommend / retrieve citable and similar cases for a source (query) case, which are well appreciated by legal experts.

\end{abstract}
\let\thefootnote\relax\footnote{This work is an extension of our SIGIR 2020 short paper~\cite{hierspcnet_sigir}. 
A part of this work was also presented
at the LDA 2019 Workshop, collocated with the International Conference on Legal Knowledge and Information Systems (JURIX) 2019; however, the workshop did not have any published proceedings. The version presented at the workshop is available at \url{https://arxiv.org/abs/2004.12307}}.

% keywords can be removed

\section{Introduction}
\label{sec:intro}
\input{sections/intro}

\section{Related Work}
\label{sec:related-work}
\input{sections/related-work}

% %\section{Problem Setting}
% %\label{sec:problem}
% %\input{sections/problem}

\section{Dataset and Experimental Setup}
\label{sec:dataset}
\input{sections/dataset}

\section{Legal Document Similarity using Network-based methods}
\label{sec:nw-mtds}
\input{sections/network-mtds}

% % \section{Adapting Text-based methods for Legal Document Similarity}
% % \label{sec:text-mtds}
% % \input{sections/text-mtds}

\section{Legal Document Similarity using Text-based methods }
\label{sec:text-mtds}
\input{sections/text-mtds}

% % \subsection{Performance analyses of the text-based methods}
% % \input{sections/results-text}

\section{Legal Document Similarity by Combining Network and Text-based methods}
\label{sec:text+nw-mtds}

\input{sections/text+network-mtds}

\subsection{Performance analyses of text and network combination methods}
\label{ss:results-txt+nw}
\input{sections/results-text+network}

% % \new{\section{Weighting similar and dissimilar document-pairs differently}}
% % \label{ss:eval-second}
% % \input{sections/eval-second}

\section{Application: Retrieving / Recommending uncited but similar case documents}
\label{sec:reco-study}
\input{sections/reco-study}

\section{Conclusion and future work}
\label{sec:conclusion}
\input{sections/conclusion}

\section*{Acknowledgements}

\noindent The authors acknowledge the anonymous reviewers whose comments greatly helped to improve the paper. 
The authors sincerely thank the Law domain experts from the Rajiv Gandhi School of Intellectual Property Law, Kharagpur, India (Amritha 
Shaji, Ankita Mohanty, and Vidisha Bhate) and from the West Bengal National University of Juridical Sciences, Kolkata, India (Dr. Shouvik Guha and Kanchan Yadav) who helped in developing the gold standard datasets. 
The research is partially supported by the TCG Centres for Research and Education in Science and Technology (CREST) through a project titled ``Smart Legal Consultant: AI-based Legal Analytics''.
P. Bhattacharya is supported by a PhD Fellowship from Tata Consultancy Services.

%Bibliography
\bibliographystyle{unsrt}  
\bibliography{references}

\end{document}

%% file: sections/intro.tex
The legal system of many countries follow the Common Law System. There are two broad types of law documents in this system -- (i)~prior case documents, also known as case laws, and (ii)~statutes, that are written laws of a particular jurisdiction, such as the Constitution of a country, laws that define and penalizes different crimes, etc.
The Common Law System gives very high importance to {\it precedents}, i.e., prior cases which are similar to a given case. 
Hence, law practitioners have to identify many (potentially, all) case documents similar to a given case, to understand and argue different legal aspects of the given case.
Given the huge number of prior cases, law practitioners/academicians require automated tools for searching and recommending similar cases for a given case.
A key task for these tools is to {\it estimate the similarity between two legal case documents}~\cite{kumar2011similarity,minocha2015finding,mandal2017measuring,hierspcnet_sigir,mandal-ailaw}. 
This is the task that we address in this paper. 

\vspace{1mm}
\noindent \textbf{The task of similarity estimation between two legal documents:}
%The task/problem considered in this work is that of \textit{similarity estimation between two legal documents}. 
The input for this task is a pair of documents (which we term as a \textit{document-pair}). 
The expected output is a \textit{single score} that reflects the level of similarity between the input document-pair.
Specifically, in this work, we consider the similarity score to be in the range $[0.0-1.0]$ where $0.0$ implies that the document-pair is least similar and $1.0$ implies that the two input documents are completely similar. 
This task is especially challenging because (i)~legal documents are long, complicated and unstructured, and (ii)~it is difficult to get large labelled datasets for training supervised Machine Learning/Deep Learning models, since annotations by legal experts are very expensive.

Note that, the methods for the similarity estimation problem that we consider in this work, have to be evaluated differently from those of a retrieval problem (where the input is a single query document and the intended output is a ranked list of `similar' documents).
%As a result, the algorithms for similarity estimation (that we develop in this work) have to be evaluated differently from the common retrieval evaluation setup.
Specifically, we evaluate the algorithms for estimating similarity between two documents as follows.
For a particular input document-pair, an algorithm generates a similarity score in [0.0-1.0].
This algorithmically generated score is then evaluated against an expert-assigned similarity score for the same document-pair in the same range [0.0-1.0]. 
Based on these two scores (one from an algorithm and another from domain experts), both in the range [0.0-1.0], the algorithm is then evaluated using measures such as Pearson Correlation and Mean Squared Error (elaborated subsequently in Section~\ref{sub:evaluation-metrics}). 

A method for similarity estimation between a pair of documents can serve as a core internal module in several applications such as prior-case retrieval / recommendation, clustering legal case documents, link prediction in the legal citation network, and so on.
In this paper, we show the applicability of our proposed similarity estimation technique in recommending / retrieving uncited case documents for a source/query document, in Section~\ref{sec:reco-study}.

\vspace{1mm}
\noindent \textbf{This work:}
Most existing methods for computing similarity between two legal case documents consider either their {\it textual content}~\cite{kumar2013similarity,mandal2017measuring}, or their {\it citation network structure}~\cite{kumar2011similarity,minocha2015finding}. 
In this work, we bring out the limitations of the existing methods, and attempt to (i)~improve existing network-based methods for legal similarity estimation, and (ii)~develop effective methods for combining signals from both the textual content and the citation network structure for improved legal document similarity estimation. 

We construct two datasets of document-pairs from the Indian Supreme Court, where the  similarities between the document-pairs are annotated by Law experts from two different reputed Law institutes in India -- (1)~the Rajiv Gandhi School of Intellectual Property Law (RGSOIPL), and (2)~the West Bengal National University of Juridical Sciences (WBNUJS).
The first dataset is used as the validation set for tuning various hyper-parameters in the similarity estimation methods, and the second dataset is used as the test set to evaluate the performance of the methods.
Using two different sets of Law experts for the validation set and test set ensures that the evaluations are not biased towards the opinions of a specific set of Law experts, and the developed methods generalize well to the similarity estimations of different Law experts.

%We evaluate our proposed approaches on case documents from the Indian Supreme Court, whose similarities have been annotated by law experts.

\vspace{1mm}
\noindent \textbf{Improving existing network-based methods for estimating similarity between legal documents}:
Existing methods for citation-network based legal document similarity~\cite{kumar2011similarity,minocha2015finding} consider a network formed from citations only to precedents/prior-case documents (which we call \underline{P}rior-case \underline{C}itation \underline{Net}work (PCNet), see Section~\ref{ss:nw-pcnet} for details). We find that citations only to precedent documents is insufficient for estimating similarity between two documents. On discussion with law experts -- three final year students and a professor from RGSOIPL, a reputed Law school in India -- we learn that knowledge about statutes is also a key aspect for understanding the similarity between case documents. Thus, existing 
network-based methods developed over PCNet miss important signals inherent in statutes.

In this work, we propose {\bf Hier-SPCNet} (Hierarchical Statute + Precedent Citation Network) that augments PCNet with a heterogeneous network of statutes. The statute network comprises of the hierarchical structure of the statutes and citations present within them.  On Hier-SPCNet, the similarity between two legal case documents is modelled as follows: {\it if two case documents cite a common precedent/statute or if two case documents cite different precedents/statutes but these precedents/statutes are themselves structurally similar in the network, then the two case documents may be based on similar legal issues. This provides an important signal for two documents being similar}. 
We develop several metapaths~\cite{dong2017metapath2vec} on Hier-SPCNet on discussion with law experts from RGSOIPL, India that capture this idea and apply metapath2vec for document similarity (we call this approach Hier-SPCNet-m2v). 
We show that by applying these domain-specific rules on Hier-SPCNet, we are able to achieve higher performance as compared to prior works that applied simplistic network based measures (e.g., bibliographic coupling, co-citation and dispersion) over PCNet. Section~\ref{ss:hierspcnet} contains details of our proposed approach.

However, Hier-SPCNet-m2v suffers from the drawback of over-estimating the similarity of certain document pairs. From our discussion with law experts -- we understand that this over-estimation is mainly because citations to all statutes/prior-cases are treated with equal importance. In the legal system of any country, there usually exists some {\it generic} statutes/prior-cases (e.g., those dealing with Equality before Law, Fundamental Rights) which are cited by a large number of cases dealing with different legal issues. However, citations to these generic statutes/prior-cases do not convey much information about legal document similarity. 

Hence, we propose a scheme that considers the discriminatory power of a node in the Hier-SPCNet network, which attenuates to a large extent the problem of over-estimation of legal document similarity in Hier-SPCNet-m2v. We propose \textit{ICF} (Inverse Citation Frequency) for statutes and Prior-case documents, which is similar to the concept of \textit{IDF} (Inverse Document Frequency) in Information Retrieval. 
We hypothesize that if two documents cite the same or similar statutes or prior-case documents, and the said statutes or prior-case documents are discriminatory enough, then the documents can be said to be similar with higher confidence. 
We materialize this hypothesis in a method named {\bf Hier-SPCNet-ICF-m2v}, where we bias the random walks of metapath2vec using ICF (see Section~\ref{ss:hspcnet-icf-m2v}). We show that similarity values inferred by Hier-SPCNet-ICF-m2v are much closer to the expert scores than the similarity varlues inferred by Hier-SPCNet-m2v (see Section~\ref{ss:results-hspcnet-m2v-icf-m2v}). 
%Through examples we demonstrate how it could attenuate as well as strengthen similarity values of some document  pairs.

% We show that an unsupervised method~\cite{mandal2017measuring} still performs better, although some of the supervised methods show promising results with scope of improvement. 

\vspace{1mm}
\noindent \textbf{Combining text-based and network-based signals}: Though both textual and network similarity individually provides important signals for estimating legal document similarity, there has not been much effort towards combining the signals from these two sources.
While there are approaches that represent the text of a document-pair as a graph~\cite{kumar2013similarity,paper2vec,article_matching}, these approaches cannot utilize the similarity signals provided by the citations to precedents and statutes (that are again connected to each other). 

To this end, we obtain text-based similarity information from a Doc2Vec model trained over legal case documents (along the lines of prior work~\cite{mandal2017measuring}), and then combine the network-based information from Hier-SPCNet with the text-based information.
We explore different techniques for combining the textual  and network-based signals for a better  estimation of legal document similarity (see Section~\ref{sec:text+nw-mtds}). We borrow various combination methods from other disciplines (e.g., methods that have been used to combine text and image embeddings), and adapt them to the task of estimating similarity between legal documents.
To the best of our knowledge, such systematic combination of the two types of signals (text-based and network-based) has not been tried earlier in the context of legal document similarity. Apart from trying simple techniques like score combination and embedding combination, we also explore state-of-the-art node embedding techniques for the purpose. 
The best performance is reported by a self-supervised neural combination method (which we call NN-Map+Conc), adopted from~\cite{nn}, which shows an improvement in correlation (with the expert-assigned similarity scores) of  11.8\% when compared to the best performing text-based method, and 20.6\% when compared to the best performing network-based method, over our test set.

\vspace{1mm}
\noindent In summary, the contributions of this work are as follows:
\begin{enumerate}
    \item We develop two datasets for the task of estimating the similarity between two legal documents, having gold standard similarity scores assigned by Law experts from two reputed Law institutes in India. 
    The datasets are available at \url{https://github.com/Law-AI/document-similarity}.
    \item We propose Hier-SPCNet, a heterogeneous network that encompasses a large body of law including both statutes and prior cases, and the citation links among them. We establish its utility for network-based legal case document similarity by incorporating domain-specific knowledge on this network. Our proposed method substantially outperforms the existing network-based similarity methods that only consider the Precedent Citation Network (consisting of case documents only).
    \item We experiment with a variety of approaches for combining textual and network information for the task of legal document similarity. In terms of correlation of the estimated similarity with expert-assigned similarity scores, our best method outperforms the state-of-the-art (SOTA) text-based method by 11.8\% and  outperforms the SOTA network-based method by 20.6\%.
    \item Finally, we show the effectiveness of our similarity estimation approach by applying it to \textit{recommend / retrieve similar documents with respect to a source/query case document}, such that the recommended documents have \textit{not} been originally cited from the source document. 
    From the ratings given by legal experts, we find that the retrieved documents are actually of much satisfaction to them -- on an average, 92\% of the top-3 most similar cases identified by our proposed method are actually considered citable by multiple law experts (see Section~\ref{sec:reco-study}). This experiment demonstrates a practical benefit of the similarity estimation method developed in this work.
\end{enumerate}

% \careful{ The added contributions of this paper are:\\
% ~(i) We improvise the method proposed in~\cite{hierspcnet_sigir} by taking into account the importance of the nodes in the network for document similarity. We add two more evaluation metrics, Mean Squared Error and F-score.\\
% ~(ii) We experiment with three supervised text-based methods for estimating document similarity.\\
% ~(iii) Towards combining text-based and network-based signals ~\cite{hierspcnet_sigir} presents only Value-based combination. In this paper we add three more families of methods.\\
% ~(iv) We show the practical utility of our proposed approach by recommending similar documents with respect to a current case document. We show that the legal experts actually find the recommended documents to be useful.}\\
The rest of the paper is organized as follows. 
Section~\ref{sec:related-work} discusses prior work on legal document similarity and related topics. 
Section~\ref{sec:dataset} describes our datasets and the experimental setup used in the work. 
Then, citation network-based methods for legal case document similarity are described in Section~\ref{sec:nw-mtds} (which also introduces the proposed Hier-SPCNet network), and 
text based methods for legal document similarity are discussed in Section~\ref{sec:text-mtds}. 
Next, Section~\ref{sec:text+nw-mtds} discusses several methods for combining network-based and text-based methods for estimating legal document similarity.
Finally, Section~\ref{sec:reco-study} applies the best performing combination method for recommending citable case documents to law experts
The study is concluded in Section~\ref{sec:conclusion}.

%% file: sections/related-work.tex
Legal data analytics, popularized through the TREC Legal Track (\url{https://trec-legal.umiacs.umd.edu/}) and the COLIEE shared task (\url{https://sites.ualberta.ca/~rabelo/COLIEE2019/}), has gained increasing popularity in recent years. 
There have been works on a variety of legal data analytics problems (see~\cite{acl-3-nlpsurvey} for a comprehensive technical survey) including question-answering in the legal domain~\cite{acm-1-qa}, statute identification given a factual description(~\cite{acm-2-cc,acm-3-cc,tkdd}), catchphrase extraction \cite{acm-4-cikm}, text classification~\cite{ipm-law-2,papaloukas-etal-2021-multi}, legal entity annotation~\cite{ipm-law-3}, legal judgement prediction~\cite{hier-bert,acl-ljp-2,ipm-law-1,ipm-law-4}, legal recommender systems~\cite{legal-reco-1,legal-reco-2,legal-cit-reco-1, legal-cit-reco-2} and so on. Pretrained transformer models on legal data~\cite{chalkidis2020legal} and its applications in several tasks is also an area of research~\cite{savelka2021discovering,icail-pretraining}.
Since annotation costs in the domain of legal data analytics is expensive, Oard et.al.~\cite{tois-1} proposed a risk minimization framework that balances the annotation costs while providing correct relevance judgements. 
Prior-case retrieval (PCR) is one of the long-standing problems in the area of legal information systems, that aims to retrieve relevant documents (noticed cases) from a pool of documents (candidate cases), given a legal fact as the query. BERT-PLI~\cite{bert-pli}, which is a state-of-the-art method for prior case retrieval, first uses a {\it legal textual entailment dataset} % (from the Federal Court of Canada) 
to train BERT for modelling similarity between paragraphs, and then uses this model to retrieve prior-cases for a query document.
Legal citation networks have also been studied for the purpose of retrieving case documents as well as scholarly legal articles~\cite{legal-cit-ir-1,legal-cit-ir-2}. 
%Especially~\cite{legal-cit-ir-2} use click rates along with citation counts. 

In this work, we consider the fundamental task of finding similarity between two legal case documents, which can be considered to be a pre-cursor to many tasks of legal information systems such as similar prior-case retrieval, legal case recommendation, and so on. 
In this section, we describe some relevant literature for the task of computing legal document similarity. We divide the existing methods into three groups 
as shown in Table~\ref{tab:relatedwork-gaps}: (i)~Text based methods, (ii)~Citation Network based methods, and (iii)~Methods for combining text and network based information. 
We separately list methods that have already been used for the task of legal document similarity, as well as some methods that have been used on generic documents (and can potentially be used on legal documents).

\begin{table}[tb]
\centering
\caption{Existing methods for document similarity, where documents may be legal as well as generic. We list only the relevant methods that have worked on text and citation network.}
\label{tab:relatedwork-gaps}
\scalebox{0.7}{
\begin{tabular}{|c|l|l|l|}
\hline
\textbf{Approach} & \multicolumn{1}{c|}{\textbf{Brief Description}} & \multicolumn{1}{c|}{\textbf{Methods applied to Legal}} & \multicolumn{1}{c|}{\textbf{Generic Methods}} \\ \hline
\begin{tabular}[c]{@{}c@{}}\textbf{Citation Network}\\ \textbf{Only}\end{tabular} & \begin{tabular}[c]{@{}l@{}}Considers a citation network of precedents / prior-cases (PCNet) ;\\ computes network measures like bibliographic coupling, co-citation \\ for similarity\end{tabular} & \begin{tabular}[c]{@{}l@{}}\textbf{Unsupervised} : PCNet~\cite{kumar2011similarity,minocha2015finding},\\This work\end{tabular} & \begin{tabular}[c]{@{}l@{}}\textbf{Unsupervised:} Uses \\ knowledge graph~\cite{kg-docsim} \end{tabular} \\ \hline
\textbf{Text Only} & \begin{tabular}[c]{@{}l@{}}Captures latent semantic information from the textual\\ content of the documents; maps the documents into vector space;\\ calculates a similarity between these vectors\end{tabular} &
 \begin{tabular}[c]{@{}l@{}}\textbf{Unsupervised:} Using Doc2Vec \\ for similarity~\cite{mandal2017measuring}\end{tabular}  & \begin{tabular}[c]{@{}l@{}} \textbf{Supervised:}  SMASH-RNN~\cite{smash_rnn},\\MaLSTM~\cite{siamese_lstm}, SMITH~\cite{smith} \end{tabular} \\ \hline
% \textbf{\begin{tabular}[c]{@{}c@{}}Constructing a Network\\ from Text\end{tabular}} & \begin{tabular}[c]{@{}l@{}}Considers textual units (sentences/paragraphs) as nodes in a graph;\\ edges exist if the textual similarity of the units are greater than a\\ threshold; computes network measures between the nodes for  similarity\end{tabular} &
% \begin{tabular}[c]{@{}l@{}} \textbf{Unsupervised:} \\ ParagraphLinks~\cite{kumar2013similarity} \end{tabular}& \begin{tabular}[c]{@{}l@{}} \textbf{Supervised:} ConceptGraph~\cite{article_matching} \\ \textbf{Unsupervised:} Paper2Vec~\cite{paper2vec} \end{tabular} \\ \hline
\begin{tabular}[c]{@{}c@{}} \textbf{Combining Text \&} \\ \textbf{Citation Network} \end{tabular} & \begin{tabular}[c]{@{}l@{}}Fusing embeddings of an entity coming from different sources\\ e.g., image, text and audio\end{tabular} & \multicolumn{1}{c|}{This work} & \multicolumn{1}{c|}{\ding{53}} \\ \hline
\end{tabular}
}
% \vspace{-5mm}
\end{table}

\subsection{Citation network-based methods for legal document similarity}
\label{ss:relatedwork-nw-mtds}

Prior works~\cite{kumar2011similarity,minocha2015finding} construct a citation network. In this network, nodes are case documents. A directed edge $d_1 \rightarrow d_2$ exists if $d_1$ cites $d_2$. We refer to this network as {\it Precedent Citation Network} (PCNet) in the rest of the paper.
On PCNet, similarity between documents $d_1$ and $d_2$ is estimated using network measures like Bibliographic coupling~\cite{kumar2011similarity}, co-citation~\cite{kumar2011similarity} and dispersion~\cite{minocha2015finding}. We use these measures as baselines in this work (details in Section~\ref{ss:nw-pcnet}).

% Network-based methods for legal document similarity consider a citation network of prior-cases. The methods construct a {\it Precedent Citation Network} (PCNet) in which the vertices are case documents, and there is a directed edge $d_1 \rightarrow d_2$ if document $d_1$ cites document $d_2$~\cite{kumar2011similarity,minocha2015finding}. 
% On PCNet, network measures such as Bibliographic coupling~\cite{kumar2011similarity}, co-citation~\cite{kumar2011similarity} and dispersion~\cite{minocha2015finding} are applied to estimate the similarity between two documents. These measures are used as baselines in this work, and explained in detail in Section~\ref{ss:nw-pcnet}.

\vspace{1mm}
\noindent \textbf{Limitations of these methods:} PCNet considers only citations to precedents/ prior-case document. The fact that a case document also contains citations to statutes, are missed in PCNet. Additionally, the similarity measures that have been tried on PCNet till date are too naive to take into account intricacies between two legal case documents. 

This work aims at improving citation-network based legal document similarity by augmenting the PCNet with a heterogeneous network of Statutes (see Section~\ref{ss:hierspcnet} for details). 

\subsection{Text based methods for document similarity}
\label{ss:relatedwork-txt-mtds}
Computing similarity between two \textit{text} documents is a generic problem, and has been applied to news documents, Wikipedia articles, question pairs, legal documents, and so on.
The task of similarity computation between two documents can be modelled both in an unsupervised as well as a supervised setup. 

\noindent \textbf{(a) Supervised methods:} 
The supervised setup labels a document-pair as 1 (similar) or 0 (dissimilar).
Large amounts of expert-annotated data is expensive to obtain in the legal domain, which makes it difficult to train supervised models for the task. Nevertheless, there have been some attempts to this end.
For instance, some works have attempted to estimate similarity in {\it triplet}s of legal documents -- given three legal case documents $(A,B,C)$, the task is to find which of $B$ or $C$ is more similar to $A$~\cite{triple-sim}. 
This task is different from the task which we focus on in this work, which is to estimate the similarity between \textit{two} legal documents.

%Prior-case retrieval models such as BERT-PLI~\cite{bert-pli} also attempt to estimate similarity between (parts of) legal documents. This method assumes the existence of a legal textual entailment dataset for training the similarity estimation models~\cite{bert-pli}. 

There exists several methods for computing long document similarity in the non-legal domain (e.g., news documents, Wikipedia articles, scientific articles). All these supervised document similarity methods consider the task as a 0/1 classification task, where 0 implies that the documents are dissimilar and 1 implies they are similar. Methods like SMASH-RNN~\cite{siamese_lstm} and SMITH~\cite{smith} use hierarchical siamese networks. There also exists graph-based approaches for the task, e.g.,~\cite{article_matching}.

While the above supervised methods can theoretically be applied for legal document similarity estimation, they need a large training set of similar / dissimilar document-pairs. Developing such datasets in the legal domain is expensive. We tried to develop such training datasets synthetically, but found that these methods do not perform well using such synthetically generated training data. Hence, we do not report these methods in this paper.

\noindent \textbf{(b) Unsupervised methods:} 
Since it is difficult to obtain large training data in the legal domain for training the supervised learning models, unsupervised methods have been mostly explored for the task of estimating the similarity between two legal documents. \cite{mandal2017measuring,mandal-ailaw} experiment with different unsupervised methods for legal document similarity. Specifically, they consider eight document representation techniques (whole document, summary, catchprases of the document and others) and seven ways to embed these representations into a vector (TF-IDF, word2vec, doc2vec, pre-trained BERT~\cite{devlin-bert}, and others). They show that Doc2Vec on the full text document gives the best performance. We adopt some of these methods for measuring textual similarity between two legal documents in this work.

\subsection{Methods that combine text and citation network}
\label{ss:relatedwork-txt+nw}

We now talk about methods that try to combine text embedding and network embedding, developed separately and independently.
Although there exists no such method that specifically tries to fuse text representation and network representation of an entity (here, a document), there exist several methods that try to learn such \textit{multi-modal embeddings} for an entity (e.g., cat) from other sources such as text (e.g., the word `cat') and image (e.g., the image of a cat). 
The Concatenation Model~\cite{conc} is a simple concatenation of normalized textual and image vectors. The Mapping Model~\cite{nn} first learns a mapping function using feed-forward neural networks from text to image modality. Then the mapping function is applied on the textual embeddings to get {\it `predicted' image vectors}. 
The normalized textual and predicted image vectors are then concatenated to obtain the multi-modal word representation. 
Stacked Auto-encoder Models~\cite{autoencoder,autoencoder2} have also been applied to learn multi-modal word representations from text and images. 

In this work, we apply these methods for unifying textual and network embeddings of legal documents. Details of these architectures are stated in Section~\ref{sec:text+nw-mtds}.

\vspace{3mm}
\noindent {\bf This paper as an extension of our prior work:} It can be noted that the idea of using a heterogeneous citation network (Hier-SPCNet) for legal document similarity was introduced in our prior work~\cite{hierspcnet_sigir} which primarily focused on network-based similarity. The present work is a much extended version of~\cite{hierspcnet_sigir}, where (i)~we improve upon the network-based method described in~\cite{hierspcnet_sigir}, and (ii)~we apply and compare among multiple methods for combining network-based similarity and text-based similarity.
Overall, the best method developed in this work significantly improves the estimation of legal document similarity, as compared to the methods presented in our prior work~\cite{hierspcnet_sigir}. In addition, we explore the practical utility of our approach in retrieving citable prior-cases given a case document.

%% file: sections/dataset.tex
In this work, we consider  documents from the Indian judiciary. This section describes the dataset and experimental setup. For a better understanding of the dataset, we give a brief background on legal documents in the Indian judiciary.

\subsection{Brief background on legal documents in the Indian judiciary}

There are two broad types of legal documents in a Common Law judiciary, such as the Indian judiciary -- {\it case documents} and {\it statutes}. 

Case documents are judgements decided in the Indian Courts (e.g. Tribunals, High Courts, Supreme Court of India). The documents contain the legal facts and issues being contended in the cases, arguments given by the parties, citations to relevant prior-case documents and statutes, and the final judgement given by the Court. An example case document can be found here~\url{https://indiankanoon.org/doc/87347014/}. 
There are two broad types of cases -- Civil cases and Criminal cases. This type is mentioned as a metadata in a case document. 

Statutes or Acts are bodies of written laws for a particular jurisdiction. Examples of Acts in the Indian judiciary include the Constitution of India 1950, Indian Penal Code 1860 (IPC), Code of Civil Procedure 1908, Code of Criminal Procedure 1973 and many others. The Acts have a hierarchical structure, with the main laws written in the leaf nodes, called {\it Articles} (used only when the leaf node is a part of the Constitution of India 1950) or {\it Sections} (for other leaf nodes, not a part of the Constitution of India 1950). Examples include Article 15 of the Constitution of India that prohibits discrimination on grounds of religion, race, caste, sex or place of birth; Section 299 of The Indian Penal Code 1860 that defines Culpable Homicide, and so on.

Now, depending on the size of the Act, there can be different hierarchical levels (or internal nodes) such as {\it Part}, {\it Chapter} and {\it Topic}.  For instance, the Part 3 of the Constitution of India 1950\footnote{Refer to \url{https://www.advocatekhoj.com/library/bareacts/constitutionofindia/index.php?Title=Constitution\%20of\%20India,\%201949} for the full Act} defines the Fundamental Rights (Articles 12 -- 35). This Part is further divided into topics `General' (Articles 12 -- 18), `Right to Freedom' (Articles 19 -- 22), `Right against Exploitation' (Articles 23 -- 24 ), `Right to Freedom of Religion' (Articles 25 -- 28), $\dots$, `Right to Constitutional Remedies' (Articles 32 -- 35).
The reader can refer to Figure~\ref{fig:act-structure} for a pictorial representation of the structure of an Act in the Indian judiciary.

The case documents cite each other and also cite the Acts/Sections. The Acts/Sections also cite each other frequently.
Thus, the relationship among the legal documents can be modeled as a heterogeneous citation network containing various types of nodes (Documents, Sections, Parts, Chapters, etc.), as detailed later in Section~\ref{ss:hierspcnet}).

\subsection{The datasets used in this work}

The evaluation set for a study on legal document similarity consists of a set of case document pairs, where each pair has to be examined by law experts to judge the similarity between the two documents. Note that, legal case documents are very long, often spanning tens to hundreds of pages. Hence even law experts need significant amount of time to understand two case documents and then judge their similarity. Also, since the similarity between two legal documents is subjective, each document-pair has to be judged by multiple law experts. As a result, developing an evaluation set for such a study is very expensive, requiring involvement of multiple law experts for long periods of time. 
This is why prior works on computing legal document similarity used relatively small datasets of $50$ document pairs~\cite{kumar2013similarity,mandal2017measuring} or even as little as $20$ document-pairs~\cite{kumar2011similarity} and $5$ document-pairs in~\cite{minocha2015finding}.

We collected $53,210$ publicly available case documents from the \emph{Supreme Court of India} and and $12,814$ {\it Acts} from the Indian judiciary.
We construct two datasets -- a validation set and a test set -- for the experiments in this paper. The validation dataset was annotated by law experts from RGSOIPL, India. On this dataset we tune the hyper-parameters of the various similarity estimation methods. A separate dataset, annotated by law experts from WBNUJS, India, was used to test the different methods. 
In particular, we got the two datasets annotated by Law experts from two different institutes, to check if the methods tuned over the validation set generalize well to the test set whose similarities are judged by different Law experts.
Table~\ref{tab:datasets} gives a brief statistics of the datasets. We now describe how these datasets were constructed.

\vspace{2mm}
\noindent {\bf Validation Set:} 
For the present study, we construct a validation set of $100$ case document-pairs. To this end, we reuse the $50$ document pairs from~\cite{mandal2017measuring}. 
The remaining $50$ document-pairs were sampled as follows --
(i)~25 document-pairs were sampled based on textual similarity, where we derive document vectors from a trained Doc2Vec model (details later in Section~\ref{sec:text-mtds}) and calculate the cosine similarity between these vectors.
(ii)~the remaining 25 document-pairs were sampled based on citation network similarity, specifically Node2Vec based similarity on prior-case citation network (detailed in Section~\ref{ss:nw-pcnet}).

We wanted the validation set to contain very similar document-pairs, moderately-similar document-pairs, and dissimilar document-pairs.
Hence, for each of the two methods, we divide the similarity values into 3 buckets $[0.0,0.4), [0.4,0.7), [0.7,1.0]$ and then pick approximately equal number of document-pairs from each bucket. 

Finally, we asked three legal experts (senior law students from RGSOIPL, India) to judge the similarity of these $100$ document-pairs independently. 
Every annotator was asked to give a score in $[0.0, 1.0]$ to every document-pair, where $0.0$ means that the document-pair is completely dissimilar and $1.0$ means that the documents are highly similar.

\vspace{3mm}
\noindent {\it Inter-annotator agreement:}
The task of document similarity estimation is inherently subjective in nature, and there can be disagreements about a particular document-pair even among domain experts.
%While we observe there was reasonably good agreement for a large majority of the document-pairs, there were disagreements among the annotators for a few document-pairs. 
Hence we measure the Inter-annotator agreement (IAA) among our expert annotators.

As detailed above, each law expert assigned a similarity score to each document-pair in the range $[0.0, 1.0]$. Consider a document-pair $(d_1,d_2)$. Assume that Expert~1 gives a similarity score $0.3$ and Expert~2 gives a similarity score of $0.2$ to this document-pair. 
Now, there are two choices for calculating the IAA -- 
(i)~in a regression setup, these scores denote a fairly high agreement between the annotators, 
(ii)~in a classification setup, if we consider each score to be a `class', then Expert~1 has assigned $(d_1,d_2)$ a `class 0.3' and Expert~2 has assigned $(d_1,d_2)$ a `class 0.2'; this implies a total disagreement between the two experts.

In our setting, we find the regression setup for calculating IAA more suitable than the Classification setup. Therefore we use \textbf{Pearson Correlation} between the expert scores as the inter-annotator agreement (IAA) measure. Since we have three annotators, we calculate Pearson correlation between the sets of scores given by each pair of annotators. We then take the average of the three correlation scores.
The inter-annotator agreement measured in the above way is $0.701$. 

For the  sake of completeness, we also attempt to calculate the IAA using \textit{Fleiss Kappa} which is meant for use when annotators assign categorical ratings to items (i.e., the classification setup).\footnote{\url{https://en.wikipedia.org/wiki/Fleiss_kappa}} 
To this end, we consider a binary classification setup and group the similarity scores into two classes -- the scores in $[0.0, 0.5)$ are considered as `class~0' and the scores in Scores $[0.5, 1.0]$ are considered as `class~1'. 
As an example, for a document-pair $(d_1,d_2)$, if the similarity scores given by three experts are $(0.2, 0.3, 0.6)$ then we consider the labels to be (class~0, class~0, class~1) respectively. 
We calculate the Fleiss Kappa on this transformed class labels. 
The Fleiss Kappa thus calculated is $0.492$. 
A Fleiss Kappa score in the range $[0.41,0.60]$ between two annotators and two classes denotes moderate agreement~\cite{fliess}. Given that we have 3 annotators and document similarity in the legal domain is inherently subjective, a Fleiss Kappa score of $0.492$ can be considered an accepted level of agreement.

\vspace{3mm}
\noindent {\it Final gold-standard similarity score for the validation set:}
For a particular document-pair, we take the mean (average) of the three experts' scores and consider it as the final expert-assigned similarity score for a pair. 
In the final validation set, we have $30$ dissimilar document-pairs with expert score in the range $[0.0, 0.4)$, $35$ moderately-similar document-pairs with expert score in $[0.4, 0.7)$ and $35$ similar document-pairs having expert-assigned similarity score in $[0.7, 1.0)$. Thus, we have a balanced validation set, having very similar, moderately similar and dissimilar document pairs.

We refer to this set of 100 document-pairs as `Validation Set' in the rest of the paper. All the hyper-parameters of all similarity estimation methods discussed henceforth are tuned on this set.

% \vspace{2mm}
% \noindent {\bf Training set:}
% The supervised similarity estimation methods applied in this work require a training dataset.
% The training dataset has been synthetically generated from the remaining case documents (leaving out the documents in the evaluation set); details are in Section~\ref{sec:text-mtds} and Section~\ref{sec:text+nw-mtds}.
\begin{table*}[]
\centering
\caption{Description of the Validation and Test datasets used in this work, with a similarity score for every document-pair (in $[0,1]$) assigned by domain experts. IAA (Inter-Annotator Agreement) is measured in terms of Pearson Correlation. Both datasets are designed to contain dissimilar document-pairs with expert score in $[0.0, 0.4)$, moderately-similar document-pairs with expert score in $[0.4, 0.7)$ and similar document-pairs having expert-assigned similarity score in $[0.7, 1.0)$.}
\label{tab:datasets}
\begin{tabular}{|c|c|c|c|c|c|c|}
\hline
\textbf{Dataset} & \textbf{Annotators} & \textbf{Dataset Size} & \textbf{IAA} & \textbf{\begin{tabular}[c]{@{}c@{}}\#pairs in\\ {[}0.0, 0.4)\end{tabular}} & \textbf{\begin{tabular}[c]{@{}c@{}}\# pairs in\\ {[}0.4, 0.7)\end{tabular}} & \textbf{\begin{tabular}[c]{@{}c@{}}\# pairs in\\ {[}0.7, 1.0)\end{tabular}} \\ \hline
Validation & RGSOIPL, India & 100 doc-pairs & 0.701 & 30 & 35 & 35 \\ \hline
Test & WBNUJS, India & 90 doc-pairs & 0.772 & 28 & 38 & 24 \\ \hline
\end{tabular}
\end{table*}

\vspace{3mm}
\noindent {\bf Test Set}: We curate another set of 90 document-pairs as the Test Set. The document-pairs were sampled in the same way as described above for the validation set  -- using Doc2Vec and Node2Vec similarity methods, and then picking equal numbers of samples from the similarity score buckets.

The document-pairs were annotated by two law experts (a research scholar and an Assistant Professor) from the West Bengal National University of Juridical Sciences (WBNUJS), India. Note that the annotators of the validation set were from a different law school (RGSOIPL, India). 
The annotation guideline were same for the annotators. 
The Inter-annotator agreement for the test set, as measured by Pearson Correlation, was 0.772. 
For the final gold-standard similarity score between a document-pair, we take the average of the two scores given the experts for this pair, similar to what was done for the validation set. 
The final test set has 28 dissimilar document-pairs with expert scores in the range [0.0, 0.4), 38 moderately-similar document-pairs with expert score in the range [0.4, 0.7) and 24 similar document-pairs with expert score in the range [0.7, 1.0), as stated in Table~\ref{tab:datasets}.

We refer to this set of 90 document-pairs as the `Test set' in the rest of the paper. The similarity estimation models with hyper-parameters tuned over the validation set are directly applied over the test set (no further tuning of hyper-parameters is carried out over the test set).

\vspace{2mm}
\noindent \textbf{How do the experts decide the similarity score of a document-pair?} 
We had a discussion with the law experts to understand how they decide the similarity score of a document-pair. All the experts agree that the task is subjective in nature, and they often follow their intuition and legal knowledge in deciding the similarity scores. 
However, there are some common indications for similarity judgment, that are followed uniformly by all experts -- for example, if the facts, the legal reasoning, and the final judgment of the Court of both the cases are similar, the experts tend to give them a very high similarity score. 
If the facts of the two cases are similar, but the final judgment differs owing to different legal questions and reasoning given by the Court, the experts give the document-pair a somewhat lower similarity score. Obviously, if the facts of the two cases are not similar, then the experts tend to give very low similarity scores.

\vspace{2mm}
\noindent \textbf{Availability of the datasets:} We share the validation set and the test set publicly at \url{https://github.com/Law-AI/document-similarity} with the hope that these datasets will promote research on legal document similarity and Law-AI in general.

\subsection{Evaluation metrics} \label{sub:evaluation-metrics}

Our validation and test sets have $n = 100$ and $n = 90$ document-pairs, respectively. 
We apply a similarity estimation method on each document-pair. Thus, for each document-pair $i$, we have two similarity scores: $y_i \in [0,1]$ which is the expert-assigned similarity score (as described above) and $\hat{y}_i \in [0,1]$ given by the method. We evaluate the performance of the method using the following three metrics: 
\begin{enumerate}
    \item \textbf{Correlation}: We compute the Pearson Correlation coefficient $\rho(y,\hat{y})$ between the expert scores $y$ and the predicted scores $\hat{y}$. This metric has been commonly used for computing legal document similarity~\cite{kumar2011similarity,mandal2017measuring,minocha2015finding}.
    
    \item \textbf{Mean Squared Error (MSE)}: MSE is computed as $\frac{1}{n} \sum_{i=1}^{n} (y_i - \hat{y_i})^2$, where $n$ is the total number of document pairs (100 in this work). 
    
    \item \textbf{F-Score}: This metric is for a binary classification setup, where a document-pair is to be classified as similar/dissimilar. To this end, we convert both the expert scores $y$ and predicted similarity scores $\hat{y}$ to a 0/1 label. If $y > 0.5$ the expert label is $1$, else $0$. Similarly, if $\hat{y} > 0.5$ the predicted label is $1$, else $0$. We then compute macro-averaged F-Score for all the $n=100$ and $n=90$ document pairs for validation and test sets respectively.
    
\end{enumerate}

%% file: sections/network-mtds.tex
% In this section we describe network-based methods for legal document similarity. We start by describing the Precedent Citation Network (PCNet) that has been used by all existing methods. Next we describe our proposed network and similarity measures on it.
This section details the methods for estimating network-based legal document similarity. We first describe the Precedent Citation Network (PCNet) which has been used by prior works. Next we describe our proposed network and similarity measures on it.

\begin{figure}
	\centering
	\includegraphics[width=4cm,height=4cm]{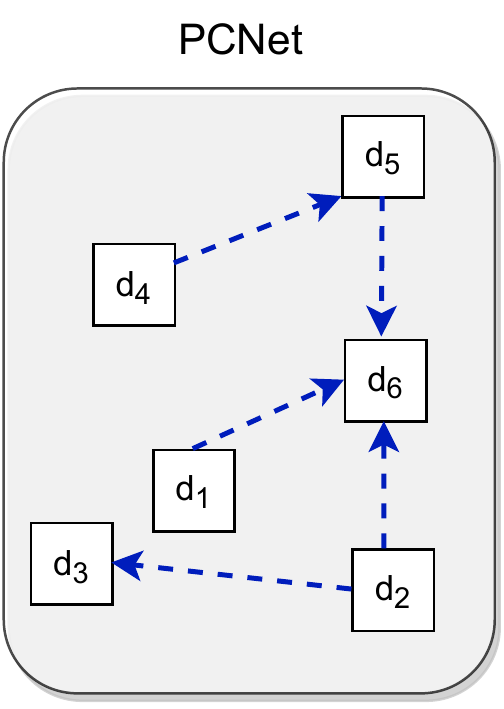}
	\caption{Precedent Citation Network (PCNet) -- the nodes are the case documents, edges exist if a document cites another.}
	\label{fig:pcnet}
\end{figure}

\subsection{PCNet and its application to Legal case document similarity}
\label{ss:nw-pcnet}

As detailed in Section~\ref{ss:relatedwork-nw-mtds}, all methods revealed by the authors' literature search for computing network-based similarity between two legal documents are based on the Precedent Citation Network (PCNet). 
Figure~\ref{fig:pcnet} shows an example PCNet.  In PCNet, the nodes of the network are the case documents. A directed edge $d_i \rightarrow d_j$ exists if document $d_i$ cites $d_j$. The following measures are then  used on PCNet to find out the similarity between two documents:

\vspace{2mm}
\noindent $\bullet$ \textbf{Bibliographic Coupling~\cite{kumar2011similarity}}: It is the fraction of overlap between the outward citations from the two documents.
For a document node $d$, let $N_{out}(d)$ be the set of out-neighbors of $d$.
For example, in Figure~\ref{fig:pcnet}, the set of out-neighbors of $d_1$ and $d_2$ are $N_{out}(d_1) = \left \{ d_6 \right \}$, and $N_{out}(d_2) =  \left \{ d_3, d_6 \right \}$.  The common out-citation is $N_{out}(d_1) \cap  N_{out}(d_2) = \left \{ d_6 \right \}$. Therefore, the similarity between ($d_1, d_2$) as measured by bibliographic coupling is $|N_{out}(d_1) \cap  N_{out}(d_2)| \; / \; |N_{out}(d_1) \cup  N_{out}(d_2)| = 1/2$.

\vspace{2mm}
\noindent $\bullet$ \textbf{Co-citation~\cite{kumar2011similarity}}: It is the fraction of overlap between the inward citations of the two documents whose similarity is to be computed. 
For a document node $d$, let $N_{in}(d)$ be the set of in-citations of $d$.
For example, in Figure~\ref{fig:pcnet}, the set of in-citations of $d_3$ and $d_6$ are $N_{in}(d_3)$ = $\left \{ d_2 \right \}$, and $N_{in}(d_6)$ =  $\left \{ d_1, d_2, d_5 \right \}$. 
The common in-citation is $N_{in}(d_3) \cap  N_{in}(d_6)= \left \{ d_2 \right \}$. Therefore, the similarity between ($d_3, d_6$) as measured by co-citation is $|N_{in}(d_3) \cap  N_{in}(d_6)| \; / \; |N_{in}(d_3) \cup  N_{in}(d_6)| = 1/3$.

\vspace{2mm}
\noindent $\bullet$ \textbf{Dispersion~\cite{minocha2015finding}}: 
Dispersion was originally used to identify social relationships in the Facebook social network~\cite{dispersion}. In the context of legal document similarity, this measure aims to
find to what extent the neighbours (out-citation documents) of two documents are themselves similar (occurs in the same community/cluster).
Consider two case documents $d_i$ and $d_j$ whose similarity is to be estimated. 
Assume that $d_i$ cites prior cases that occur in certain communities/clusters, where each community/cluster deals with a specific legal area. 
Dispersion aims to quantify to what extent $d_j$ cites prior cases that belong to the same community/cluster (legal area) whose documents are cited by $d_i$.

Mathematically, dispersion was defined in~\cite{dispersion} as follows -- the dispersion between two nodes $u,v$ is
$disp(u,v) = \sum_{s,t \in C_{uv}} d(s,t)$ where $C_{uv}$ is the set of all common neighbours between $u$ and $v$ and $d(s,t)$ is the distance function between all pairs of nodes $(s,t)$ in $C_{uv}$. As shown in~\cite{dispersion} the distance function that works best was equal to 1 when $s$ and $t$ are not directly linked and also have no common neighbors and equal to 0 otherwise.
We use the \textit{NetworkX} implementation\footnote{\url{https://networkx.github.io/documentation/networkx-1.9/reference/generated/networkx.algorithms.centrality.dispersion.html}} to compute Dispersion.

% We use the \textit{NetworkX} implementation for this measure.\footnote{\url{https://networkx.github.io/documentation/networkx-1.9/reference/generated/networkx.algorithms.centrality.dispersion.html}}

\begin{figure}
	\centering
	\includegraphics[width=8.5cm,height=5.5cm]{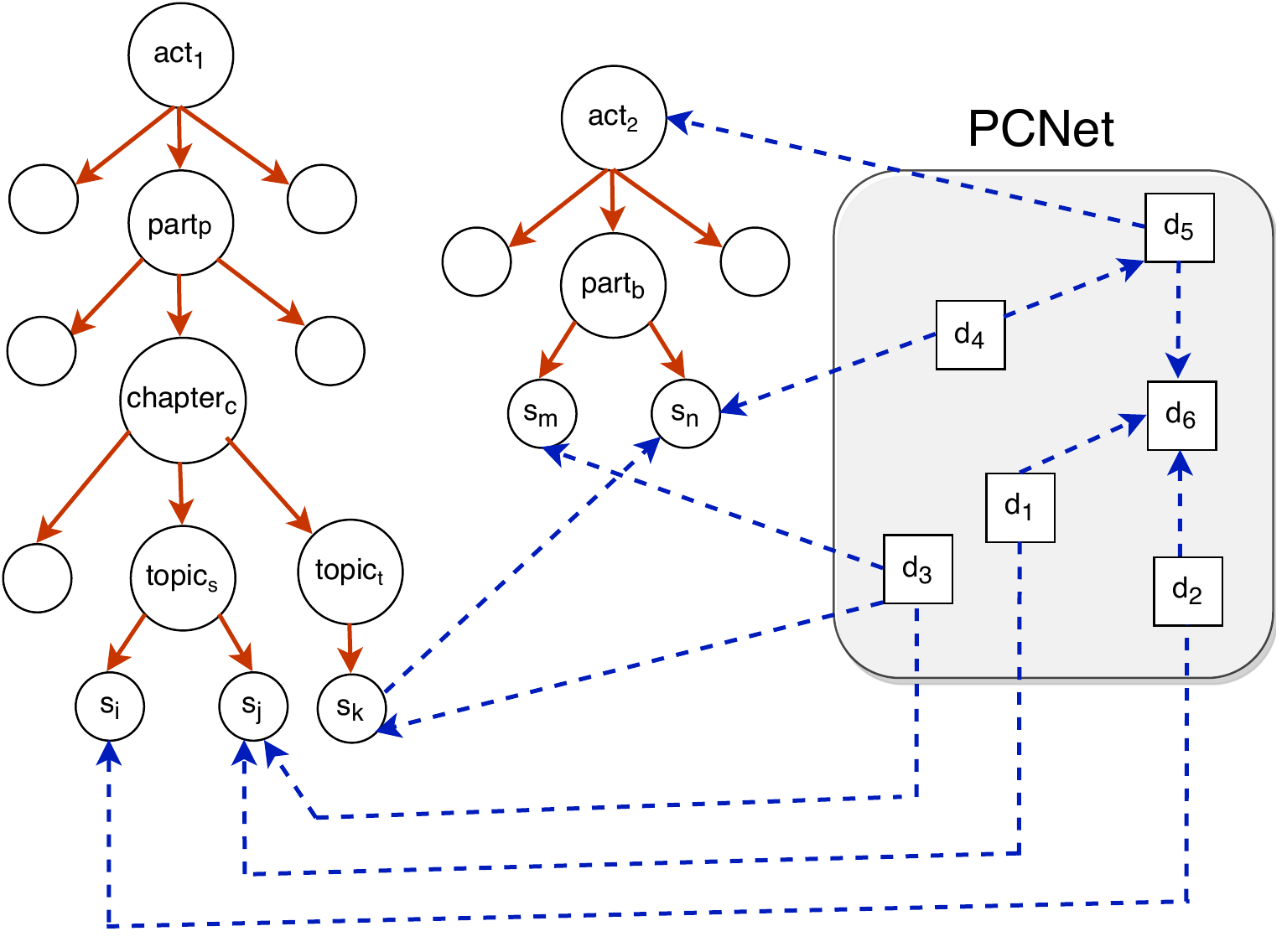}
	\caption{The proposed Hierarchical Statute+Precedent Citation Network (Hier-SPCNet). It contains case documents and statutes. The greyed box exhibits PCNet, which has been used by prior works. Figure reproduced from our prior work~\cite{hierspcnet_sigir}. }
	\label{fig:act-structure}	
	\vspace{-5mm}
\end{figure}

\subsection{The Hier-SPCNet citation network}
\label{ss:hierspcnet}

PCNet considers only case documents, and thus misses an important signal of legal similarity that is inherent in the {\it statutes} that are written laws of a particular jurisdiction (e.g., a country). 
From our discussions with law experts (from the Rajiv Gandhi School of Intellectual Property Law, India) we understand that statutes are a rich source of legal domain knowledge and is an inevitable signal for finding the similarity between two legal documents. To exploit the signals of legal similarity inherent in statutes, we propose to augment PCNet with a network of statutes. 

The resulting hierarchical, heterogeneous network is termed as Hierarchical Statute + Precedent Citation Network ({\bf Hier-SPCNet}), depicted in Figure~\ref{fig:act-structure}. Hier-SPCNet is based on the key idea that similar documents will have topologically similar neighbourhoods, and these neighbourhoods are defined both by statutes and precedents.
We now describe the Hier-SPCNet network in detail and how we construct it. % to compute the similarity between two legal documents.

\vspace{2mm}
% \noindent {\bf (I) Modeling the hierarchy of statutes:} In many common law countries, an `Act' has its own hierarchy. For instance, in the Indian judiciary, an Act can be divided into `parts'; each `part' can be divided into `chapters'; each `chapter' can be further divided into `topics'; under a `topic' are finally `sections'/`articles'. 
% An example of the Act $\rightarrow$ Part $\rightarrow$ Chapter $\rightarrow$ Topic $\rightarrow$ Section/Article hierarchy is --
% {\it Constitution of India, 1950 $\rightarrow$ Part VI: The States $\rightarrow$ Chapter III: The State Legislature $\rightarrow$ Topic: Disqualification of members $\rightarrow$ Section 192: Decision on questions as to disqualification of members.}
% Sometimes, for smaller acts, parts of this hierarchy may not be explicitly specified. For instance, we may have sections/articles directly under an act. An example is -- {\it Dowry Prohibition Act, 1961 $\rightarrow$ Section 3:  Penalty for giving or taking dowry.}

% For construction of Hier-SPCNet, we extract the hierarchy from the text of the statutes, and then represent each act as a hierarchical structure of nodes (act / parts / chapters / topics / sections) and hierarchy links. 
% Figure~\ref{fig:act-structure} shows a pictorial representation of an act having the complete hierarchy ($act_1$) and another act having a smaller hierarchy ($act_2$). 
\noindent {\bf (I) Modeling the hierarchy of statutes:} In many countries following the common law system, an `Act' has a hierarchy. For instance, in the Indian judiciary, an `Act' may be divided into several `parts'; `parts' are further divided into `chapters' which are further sub-divided into `topics'; a `topic' is further branched into `sections' / `articles'. An example of the hierarchy is --  {\it Constitution of India, 1950 $\rightarrow$ Part V: The Union $\rightarrow$ Chapter II: Parliament $\rightarrow$ Topic: Legislative Procedure $\rightarrow$ Article 107: Provisions as to introduction and passing of Bills}. The hierarchy is not uniform across all Acts. For instance, the Indian Penal Code, 1860 (IPC) has the following structure : {\it IPC $\rightarrow$ Chapter $\rightarrow$ Section}.

For creating the hierarchical structure of the statutes, we scrape various online resources and store it in a tree-like structure, having hierarchical links. Figure~\ref{fig:act-structure} shows a diagrammatic representation -- $act_1$ has a deeper hierarchy (e.g., Constitution of India, 1950) and $act_2$ has a shallow hierarchy (e.g. Indian Penal Code, 1860).

\vspace{2mm}
\noindent {\it Challenges in modeling the hierarchy of statutes}: In the absence of a uniform platform providing all statutes in the Indian judiciary, we had to identify several sources on the internet (e.g. \url{https://www.advocatekhoj.com/library/bareacts/}, \url{http://www.bareactslive.com/indexCA.html}, \url{https://indiankanoon.org/}, etc.) that provide such information. Unifying information from all these websites into a standardized format (to form the hierarchy) was a challenging task. Another related challenge was to automatically extract the hierarchy among Acts, Chapters, Topics, Sections, etc. from the text of the statute documents.

\vspace{2mm}
\noindent {\bf (II) Extraction of statute citations from text:}
Statutes are cited within the text of case documents as well as within other statutes. Precedents (case documents) are cited within text of other case documents. 
Citation extraction from legal texts is challenging because statutes/precedents citations are not systematic and there are various forms of such citations.

In this work, we design several regular expressions to automatically extract statute citations from the text of a case document or a statute. For this, we consulted Law experts as well as manually observed a large number of statute citations. 
We had to follow an iterative process to gradually improve the coverage of the set of regular expressions. 
Finally, we designed a set of regular expressions that broadly capture the following pattern -- `Section(s) / Article(s) \textit{<sequence of numbers>} of the <Act>', where the \textit{<sequence of numbers>} are observed to be written in the following ways: 
(i)~a single number, e.g.. `Section 302 of the Indian Penal Code, 1860', 
(ii)~two numbers mm and nn, e.g.. `Articles 19 and 22 of the Constitution', 
(iii)~three numbers mm, nn and pp, e.g., `Sections 14, 15 and 20 of the Income-tax Act, 1961;, and
(iv)~a range mm to nn, e.g., `Sections 50 to 55 of the Customs Act, 1962'.

The text of a case document or a statute may contain reference to a specific section of an Act, or to an Act as a whole, without mentioning any section number, e.g., ``taking dowry is a punishable offence as described in the Dowry Prohibition Act, 1962.'' 
%We consider this as a citation between a document and the Act.

\vspace{2mm}
\noindent \textit{Performance of the statute citation extraction method:} We conducted a manual evaluation of the methodology for extracting statute citations by law experts on a small randomly-selected set of $20$ documents. The law experts manually listed the set of all statute citations that are actually present in these 20 case documents. 
We observed that the heuristic-based method achieved a precision of $1.0$ implying that all the statute citations that the method captured were correct.
The method could extract correctly 90\% of all the statutes that were cited, thus giving a recall of $0.9$. The reasons why our method missed some of the citations are listed below.

\vspace{2mm}
\noindent \textit{Limitations of the statute citation extraction method:} 
We observed a few limitations of this method of identifying statute citations, due to which it missed around 10\% of the statute citations (as described above). Some of the limitations are as follows. 
(i)~Sometimes, Acts are cited without mentioning the year. There are Acts that got amended multiple times and citing such Acts without the particular year results in ambiguity e.g., Appropriation (No. 5) Act, 1964; Appropriation (No. 5) Act, 2010 ; Appropriation (No. 5) Act, 2015 etc. 
We do not consider citations to such Acts that have been amended and a citation to a version of that Act has been made by a document without specifying the particular year. 
(ii)~The method was not able to handle co-references. For example, consider a scenario where the following line has been made towards the starting of a case document -- ``an FIR was filed under the Section 11 of the Dowry Prohibition Act, 1961 (hereinafter referred to as ``the Act'').'' -- and at another part of the document, we find the citation as ``Section 7 of the said Act''. While our regular expression-based method could extract Section 11 of the Dowry Prohibition Act, 1961, it missed the citation to Section 7 of the Dowry Prohibition Act, 1961. 
%However, such type of citations were not very frequent in our dataset.
For all these reasons, the method could not capture about 10\% of the statute citations.

%\new{(iii)~We were also unable to extract a few citations that were complicated in nature. For instance, patterns of the form Section (s) / Articles(a) <sequence of numbers> of the <$Act_1$> read with Section (s) / Articles(a) of <$Act_2$> could not be identified by our method.}

% Our heuristic-based approach could not capture the citation to Section 10 of the Tamil Nadu Buildings (Lease and Rent Control) Act, 1960 in such cases, though the citation to Section 11 of the Act will be captured correctly. However, such co-references are very rarely encountered in our dataset.

\vspace{2mm}
\noindent {\bf (III) The Hier-SPCNet network:} The Hier-SPCNet network has \textbf{six kinds of nodes} -- the case documents, acts, parts, chapters, topics, sections (or articles). The network has \textbf{two types of edges} -- {\it hierarchy edges} (indicated in orange, solid lines in Figure~\ref{fig:act-structure}) and {\it citation links} (blue, dotted lines in Figure~\ref{fig:act-structure}). We describe the edge types next.

\vspace{1mm}
\noindent \textbf{Citation edges:} As the name suggests, these edges denote \textit{citation} between two vertices. The citation edges are of three types:
\textit{(1)~ document $\rightarrow$ document}: when a case document cites another case document. These edges form the PCNet (the grey coloured box in Figure~\ref{fig:act-structure}). Prior-works have used only this type of edge for network-based document similarity.
\textit{(2)~document $\rightarrow$ statute}: when a case document refers/cites a statute (a section or an article). In Figure~\ref{fig:act-structure}, $d_1$ cites section $s_i$ of $act_1$. Hence, there is an edge of this type.
\textit{(3)~document $\rightarrow$ act}: when a case document cites an Act (as a unit, without referring to a particular section of the Act), e.g., document $d_5$ cites $act_2$.
\textit{(4)~statute $\rightarrow$ statute}: there are citations between different sections of the Acts as well. A citation edge exists if a statute cites another statute. The two statutes/sections can be part of the same Act or different Acts. From Figure~\ref{fig:act-structure}, statute $s_k$ of $act_1$ cites statute $s_n$ of $act_2$.

\vspace{1mm}
\noindent  \textbf{Hierarchy edges:} These edges, indicated in orange, solid arrows in Fig.~\ref{fig:act-structure}, represent the hierarchical structure within each Act, as described earlier.
These edges are of the following types:
%such as {\it act $\rightarrow$ part} (e.g., $part_p$ is under $act_1$ in Fig.~\ref{fig:act-structure}), {\it act $\rightarrow$ chapter}, {\it part $\rightarrow$ section} (e.g., in $act_2$, sections $s_m$ and $s_n$ are under a $part_b$), 
%\textit{topic $\rightarrow$ section} (e.g., $s_i$ and $s_j$ are under $topic_s$ under $act_1$), and so on. 
We report here an exhaustive set of all types of hierarchy edges observed in the Indian legal statutes.
\begin{itemize}
		\item \textbf{act $\rightarrow$ part}: In Figure~\ref{fig:act-structure}, $part_p$ is under $act_1$, hence $act_1$ and $part_p$ are connected by an hierarchy edge.
		\item \textbf{act $\rightarrow$ chapter} : A chapter is under an act
		\item \textbf{act $\rightarrow$ topic} :  A topic is under an act.
		\item \textbf{act $\rightarrow$ section} :  A statute is under an act.
		\item \textbf{part $\rightarrow$ chapter} :  A chapter is under a part of an act. Eg. in $act_1$ , $part_p$ is connected to $chapter_c$
		\item \textbf{part $\rightarrow$ topic} :  A topic is under a part of an act. 
		\item \textbf{part $\rightarrow$ section} :  A section is under a part of an act. Eg. in $act_2$ , $s_m$ and $s_n$ are under a $part_b$.
		\item \textbf{chapter $\rightarrow$ topic} :  A topic is under a chapter of an act. Eg. in $act_1$, $topic_s$ and $topic_t$ is connected to $chapter_c$
		\item \textbf{chapter $\rightarrow$ section} :  A section is under a chapter of an act. 
		\item \textbf{topic $\rightarrow$ section} :  A section is under a topic of an act. Eg. in $act_1$, $s_i$ and $s_j$ are under $topic_s$, $s_k$ is under $topic_t$.
	\end{itemize}

As mentioned earlier, the hierarchy levels may not be uniform across all Acts.

\vspace{2mm}
\noindent \textbf{Text in Hier-SPCNet:} The document nodes (e.g., $d_1$, $d_2$, $\dots$ , $d_6$ in Fig.~\ref{fig:act-structure}) and the leaf nodes in the Statute hierarchy (e.g., $s_i$, $s_j$, $s_k$, $s_m$, $s_n$ in Fig.~\ref{fig:act-structure}) have associated textual content. 
In this section, we are focusing only on the network structure. But we shall use these textual content later in Section~\ref{sec:text-mtds} and Section~\ref{sec:text+nw-mtds}.

\vspace{2mm}
\noindent \textbf{Size of Hier-SPCNet:} From our collection of $53,211$ case documents and $12,814$ Acts (as stated earlier in Section \ref{sec:dataset}), we could extract statute and precedent citations from $30,056$ documents. 
Thus for our experiments, Hier-SPCNet consists of $124,104$ nodes ($30,056$ document nodes and the rest statute nodes) and $337,136$ edges.

\subsection{Document similarity using Hier-SPCNet and PCNet}
\label{ss:hier-spcnet-m2v}

In this section we show how Hier-SPCNet can be used for computing network-based similarity between two case documents. Measures that have been employed in prior-works on PCNet -- bibliographic coupling, co-citation and dispersion (described in Section~\ref{ss:nw-pcnet}) -- can be applied over Hier-SPCNet as well. When these measures are applied over Hier-SPCNet, they also incorporate statute information. For instance, when computing bibliographic coupling between two case documents on the Hier-SPCNet network, it considers the number of common citations to statutes and prior-cases.

We also leverage node embedding algorithms Node2Vec~\cite{node2vec} and Metapath2Vec~\cite{dong2017metapath2vec} on Hier-SPCNet and PCNet. 
Node embedding algorithms map nodes of a graph to a n-dimensional space in a way that nodes having similar neighbourhoods in the network have similar representations (embeddings). cosine-similarity between the two node (document) embeddings, renders the similarity between two case documents.

\vspace{2mm}
\noindent \textbf{Node2Vec~\cite{node2vec}}:
Given a network, Node2Vec considers it to be homogeneous (i.e. nodes are assumed to be of the same type). 
The node embeddings (vectors) are generated via random walks of a given length (a hyperparameter) over the network, following Breadth First Search (BFS) or Depth First Search (DFS).  As an analogy to word2vec, each random walk is a sentence, with words being the vertices in the random walk. The total number of random walks (hyper-parameter) is the total number of sentences in the corpus. Thus, having converted a graph to a corpus, it then gets the vectors of each word (vertex) using the skip-gram architecture of word2vec.

Node2Vec is applied on both PCNet and Hier-SPCNet. We used the publicly available implementation\footnote{\url{https://github.com/aditya-grover/node2vec}}. The node embedding size was $200$ and other hyperparamater values were default.
Node2Vec considers the network to be homogeneous, i.e., all the vertices and edges to be having uniform semantics. Note that, while PCNet is actually homogeneous, Hier-SPCNet is heterogeneous. When Node2vec is applied on Hier-SPCNet, it considers the different vertices and edges to be of the same type.
% Note that Node2vec assumes a network to be homogeneous (all nodes and edges of same type). While PCNet is actually homogeneous, Hier-SPCNet is not; however, Hier-SPCNet is also considered homogeneous when applying Node2vec. In this case, both the document and statue nodes are implicitly considered to be of the same type.

\vspace{2mm}
\noindent \textbf{Metapath2Vec~\cite{dong2017metapath2vec}}: 
In Metapath2Vec, the input network is considered to be heterogeneous -- the vertices and edges are of many types, each having different semantics. Metapath2Vec follows a similar algorithm like Node2Vec, the difference being that while Node2Vec uses Breadth First/Depth First search, Metapath2Vec operates on {\it metapaths} defined  by the user. 

% A metapath is a path that exists between two vertices where the edges along the path can have different semantics. 

Formally a metapath is defined as a sequence of relations defined between different object types. Consider a heterogeneous graph (here, the Hier-SPCNet graph) as $G = (V,E,T)$, where $V$ is the set of nodes, $E$ is the set of edges, $T$ denotes the different node types (e.g., section, topic, chapter, part, act, document) and $t \in T$. A metapath $P = V_1 \overset{R_1}{\rightarrow} V_2 \overset{R_2}{\rightarrow} \cdots V_t \overset{R_t}{\rightarrow} V_{t+1} \cdots \overset{R_{l-1}}{\rightarrow} V_{l}$ defines a composite relation $R = R1 \circ R2 \circ \dots \circ R_{l-1}$ between node types $V_1$ and $V_l$, where $\circ$ denotes the composition operator on relations (e.g., $doc \rightarrow sec \rightarrow act \rightarrow sec \rightarrow doc$).

Based on discussions with our law experts from RGSOIPL (who annotated the validation set), we form the following hypothesis -- {\it if two case documents cite a common precedent/statute, or if two case documents cite different precedents/statutes but these precedents/statutes are themselves structurally similar in the network, then the two case documents may be based on similar legal issues. This provides an important signal for two documents being similar}.
We obtain detailed suggestions from the law experts as to what type of citations from two case documents can give a good estimate of the similarity between them. 
We encode this domain knowledge by defining {\bf 14 different metapaths} over Hier-SPCNet that cover the different ways by which two documents can cite the same or related statutes/prior-cases. Through this we expect to capture a notion of similarity between the case documents.
%The structural similarities are defined by the metapaths. 
In every metapath, the source and target nodes of the path are documents. 
Note that, while Hier-SPCNet is a directed graph, the metapaths being a sequence of node types are essentially undirected paths.
Specifically, we define the following metapaths for computing similarity between document nodes in Hier-SPCNet:
\begin{itemize}
    \item \textbf{doc-sec-doc}: this metapath captures the situation where the same section/article of an Act is cited by both the case documents. For example, in Figure~\ref{fig:act-structure}, both the case documents $d_1$ and $d_3$ cite the section $s_j$. 
    
    \item \textbf{doc-sec-topic-sec-doc}: when different sections/articles are cited by the two case documents, but these sections/articles are under the same `topic'. For instance, in Fig.~\ref{fig:act-structure}, $s_j$ is cited by $d_1$, $s_i$ is cited by $d_2$ and both $s_i$ and $s_j$ are under the same topic $topic_s$.
    Similarly, there is \textbf{doc-sec-part-sec-doc} (E.g., $s_m$ is cited by $d_3$, $s_n$ is cited by $d_4$ and $s_m$ and $s_n$ are under $part_b$ of $act_2$.), \textbf{doc-sec-chapter-sec-doc} and \textbf{doc-sec-act-sec-doc}.
    
    \item \textbf{doc-sec-topic-act-topic-sec-doc} :  when different sections/articles, under different `topics' are cited by the two case documents, but the `topics' are under the same `act'.
    Similarly, we have 
    \textbf{doc-sec-chapter-act-chapter-sec-doc} (sections/articles under different chapters) and \textbf{doc-sec-part-act-part-sec-doc} (sections/articles under different parts.) 
    
    \item \textbf{doc-sec-topic-chap-topic-sec-doc}: when different sections are cited by the two documents but the sections belong to the same chapter. For example, in Fig.~\ref{fig:act-structure}, $s_j$ is cited by $d_1$, $s_k$ is cited by $d_3$ and $s_i$ and $s_k$ belong to different topics but under the same $chapter_c$ of $act_1$.
    
    \item \textbf{doc-sec-chapter-part-chapter-sec-doc}: when different sections are cited by the two case documents and the sections belong to different chapters but under the same part.
    
    \item \textbf{doc-sec-topic-part-topic-sec-doc}:  when different sections are cited by the two case documents and the sections belong to different topics but under the same part.

    \item \textbf{doc-sec-sec-doc}: when different sections of different acts are cited by two case document, but one of the sections cites the other section.
    
    \item \textbf{doc-act-act-doc}: when different acts  (as a unit) are cited by the two case documents and  one of the acts cite the other act.

    \item \textbf{doc-doc-doc}: when a common precedent/prior-case is cited by the two case documents. Note that this is the only metapath that could be applied over PCNet.
\end{itemize}

\vspace{3mm}
\noindent \underline{Implementation details:}
For Metapath2Vec, we used the publicly available implementation\footnote{\url{https://stellargraph.readthedocs.io/en/stable/demos/embeddings/metapath2vec-embeddings.html}}, with walk length set to 7 (because the longest metapaths have a sequence of 7 nodes).
We experimented with values 500, 1000, 2000 and 3000 for the hyperparameter `number of random walks per root node' and found 2000 to be giving the best result on the validation set. 
Hence we report all results considering 2000 random walks per root node.
We chose an embedding size of 200 to keep the textual embedding (described later in the paper) and network embedding sizes equal. 
Other hyper-parameters set to default in the publicly available implementation stated earlier.

\vspace{3mm}
\noindent After applying Metapath2vec (with the above metapaths) to Hier-SPCNet, we obtain an embedding of size $200$ for every node. The similarity between two case documents is computed as the cosine similarity between the embeddings of the two corresponding nodes.
We refer to this method of inferring document similarity as \textbf{Hier-SPCNet-m2v} in the rest of the paper.

\subsection{Examples to show the working of metapaths on Hier-SPCNet}
\label{ss:hspcnet-m2v-egs}

\begin{table}[]
\caption{A working example about how the similarities between the two document pairs (1985\_113, 1991\_12) and (1987\_189, 1991\_48) are inferred by Hier-SPCNet-m2v (for Statutes, x\_n implies Section n of act x)}. 
\label{tab:m2v-working-example}
\scalebox{0.5}{
\begin{tabular}{lllll}
\hline
\multicolumn{1}{|c|}{\textbf{\begin{tabular}[c]{@{}c@{}}Docs in\\ the Pair\end{tabular}}} & \multicolumn{1}{c|}{\textbf{Statutes Cited}} & \multicolumn{1}{c|}{\textbf{Precedents Cited}} & \multicolumn{1}{c|}{\textbf{Relevant Metapaths}} & \multicolumn{1}{c|}{\textbf{\begin{tabular}[c]{@{}c@{}}Demonstration of how the metapaths capture\\ the similarity between the documents\end{tabular}}} \\ \hline
\multicolumn{5}{|c|}{\textit{Expert Similarity Score = 0.87 ; Network Similarity inferred by Hier-SPCNet-m2v  = 0.579}} \\ \hline
\multicolumn{1}{|l|}{1985\_113} & \multicolumn{1}{l|}{\begin{tabular}[c]{@{}l@{}}National Security Act 1980 (NSA)\_3, NSA\_10\\ NSA\_11, NSA\_12, Constitution\_21, Constitution\_226 ;\\ Conservation of Foreign Exchange and Prevention of \\ Smuggling Activities Act (COFEPOSA)\_8\end{tabular}} & \multicolumn{1}{l|}{1969\_324 ; 1981\_T\_2} & \multicolumn{1}{l|}{doc-sec-doc} & \multicolumn{1}{l|}{1991\_12 -- COFEPOSA\_8 -- 1985\_113} \\ \hline
\multicolumn{1}{|l|}{\multirow{3}{*}{1991\_12}} & \multicolumn{1}{l|}{\multirow{3}{*}{\begin{tabular}[c]{@{}l@{}}Preventive Detention Act 1950\_3 ; Constitution\_22 ; \\ General Clauses Act 1897\_21; COFEPOSA\_3, COFEPOSA\_8 \end{tabular}}} & \multicolumn{1}{l|}{\multirow{3}{*}{\begin{tabular}[c]{@{}l@{}}1979\_P\_21 ; 1969\_S\_236 ; 1989\_U\_69 ; \\ 1989\_V\_7 ; 1969\_C\_47 ; 1971\_B\_3 ; \\ 1969\_324 ; 1974\_H\_7 ; 1974\_K\_27 ; \\ 1980\_39 ; 1969\_S\_116 ; 1975\_J\_27\end{tabular}}} & \multicolumn{1}{l|}{doc-sec-topic-sec-doc} & \multicolumn{1}{l|}{\begin{tabular}[c]{@{}l@{}}1991\_12 -- Constitution\_22 -- Constitution\_Topic 2 --\\ Constitution\_21 -- 1985\_113\end{tabular}} \\ \cline{4-5} 
\multicolumn{1}{|l|}{} & \multicolumn{1}{l|}{} & \multicolumn{1}{l|}{} & \multicolumn{1}{l|}{doc-sec-act-sec-doc} & \multicolumn{1}{l|}{\begin{tabular}[c]{@{}l@{}}1991\_12 -- COFEPOSA 1974\_3 -- COFEPOSA 1974 -- \\ COFEPOSA 1974\_8 -- 1985\_113\end{tabular}} \\ \cline{4-5} 
\multicolumn{1}{|l|}{} & \multicolumn{1}{l|}{} & \multicolumn{1}{l|}{} & \multicolumn{1}{l|}{doc-doc-doc} & \multicolumn{1}{l|}{1991\_12 -- 1969\_324 -- 1985\_113} \\ \hline
\multicolumn{5}{|c|}{\textit{Expert Similarity Score = 0.73 ; Network Similarity inferred by Hier-SPCNet-m2v = 0.681}} \\ \hline
\multicolumn{1}{|l|}{1987\_189} & \multicolumn{1}{l|}{Constitution\_22 ; NSA\_3} & \multicolumn{1}{l|}{\begin{tabular}[c]{@{}l@{}}1969\_S\_236 ; 1978\_B\_1 ; \\ 1981\_S\_270 ; 1973\_S\_220 ;\\ 1972\_S\_154 ; 1969\_S\_116\end{tabular}} & \multicolumn{1}{l|}{doc-sec-doc} & \multicolumn{1}{l|}{1991\_48 -- Constitution\_22 -- 1987\_189} \\ \hline
\multicolumn{1}{|l|}{\multirow{1}{*}{1991\_48}} & \multicolumn{1}{l|}{\multirow{1}{*}{\begin{tabular}[c]{@{}l@{}}Constitution\_32, Constitution\_22, Prevention of Illicit Traffic\\ in Narcotic Drugs and Psychotropic Substances Act 1988 (NDP)\_2,\\ NDP\_3, NDP\_9, NDP\_10\end{tabular}}} & \multicolumn{1}{l|}{\multirow{2}{*}{\begin{tabular}[c]{@{}l@{}}1969\_S\_236 ; 1991\_12 ; 1969\_C\_47 ; \\ 1971\_B\_3 ; 1974\_H\_7 ; 1969\_S\_116 ; \\ 1975\_J\_27\end{tabular}}} & \multicolumn{1}{l|}{doc-sec-sec-doc} & \multicolumn{1}{l|}{\begin{tabular}[c]{@{}l@{}}1987\_189 -- NDP\_3 -- constitution\_22 -- 1991\_48\end{tabular}} \\ \cline{4-5} 
\multicolumn{1}{|l|}{} & \multicolumn{1}{l|}{} & \multicolumn{1}{l|}{} & \multicolumn{1}{l|}{doc-doc-doc} & \multicolumn{1}{l|}{\begin{tabular}[c]{@{}l@{}}1987\_189 -- 1969\_S\_236 -- 1991\_48\\ 1987\_189 -- 1969\_S\_116 -- 1991\_48\end{tabular}} \\ \hline
\end{tabular}
}
\vspace{-5mm}
\end{table}

Table \ref{tab:m2v-working-example} shows examples of two document-pairs. In the ``Statutes Cited'' column we show the statutes cited in the corresponding document. The next column states the precedents cited by each document. 
We show some relevant metapaths applicable for a document-pair in the fourth column. Finally in the last column, we demonstrate how a certain metapath materializes in a document pair.

For the first pair (1985\_113 and 1991\_12)\footnote{These documents can be accessed at \url{http://www.liiofindia.org/in/cases/cen/INSC/1985/113.html} and \url{http://www.liiofindia.org/in/cases/cen/INSC/1991/12.html} respectively.}, the similarity inferred by metapath2vec is $0.579$ and the average expert similarity score is $0.87$. To understand the reason behind this, we trace the metapath that connects these two documents. 
There are four such metapaths (i)~doc-sec-doc -- implies that the documents cite the same section -- document 1985\_113 cites section 8 of the COFEPOSA Act, which is also cited by 1991\_12, 
(ii)~doc-sec-topic-sec-doc -- the documents cite sections under the same topic of an act -- 1991\_12 cites Article 22 of the Constitution which comes under Topic 2 of the Act (Constitution). Under this Topic 2 is also Article 21 which is cited by 1985\_113, the other document of the pair. 
(iii)~doc-sec-act-sec-doc -- the documents cite sections under the same act, and 
(iv)~doc-doc-doc -- implies that there is a common precedent/prior-case document (1969\_324) cited by both documents.

For the second pair (1987\_189 and 1991\_48)\footnote{These documents can be accessed at \url{http://www.liiofindia.org/in/cases/cen/INSC/1987/189.html} and \url{http://www.liiofindia.org/in/cases/cen/INSC/1991/48.html}}., we find that metapaths doc-sec-doc, doc-sec-sec-doc and doc-doc-doc hold. 
The metapath doc-sec-doc captures the common statutes that have been cited by the documents. The metapath doc-sec-sec-doc captures a non-trivial aspect -- the two documents cites different statutes that come under different Acts (i.e., the cited statutes are not in the hierarchy of the same Act), but there exists a citation link between the two statutes. This is shown by Section 3 of the NDP Act and Article 22 of the Constitution where the former cites the latter. 

Through these examples, we show that, through the metapaths we are able to assemble different signals of legal document similarity, which are otherwise difficult to be captured through simplistic measures like bibliographic coupling. 
These matapaths effectively encode the legal domain knowledge and is able to explain to some extent the similarity between two legal documents. Apart from the performance benefits (detailed in the next section), the metapaths also contribute to explainability (which is a very important aspect in legal analytics) of why two documents are inferred to be similar.

\subsection{Results of the network-based methods on PCNet and Hier-SPCNet}
\label{ss:results-pcnet-hspcnetm2v}

\input{sections/results-network-1}

\subsection{Improving the similarity estimation -- accounting for the discriminatory power of nodes}
\label{ss:hspcnet-icf-m2v}

%From our experiments, we observed that the Hier-SPCNet-m2v method often {\it over-estimates} the similarity between case-pairs. 
From our discussions with legal experts, we understand that in many legal jurisdiction, there are some {\it generic} statutes and prior-cases which are cited by many cases from various domains of law. 
Though these statutes/prior-cases are very important (which is why they are cited by many cases), citation links to these generic statutes/cases lack the ability to discriminate between different domains of law.

The top 3 most frequently cited statutes in our dataset from Indian judiciary are listed in Table~\ref{tab:icf-eg}. The first column is the name of the statute, next is the title and in the third column are few categories of case documents that cite these statutes (the categories were understood manually; the case documents in Indian judiciary do not specify any category). 
For instance, a frequently cited statute is Section 302 of the Indian Penal Code (IPC) which is about Punishment for Murder. 
Our legal experts explain that murder can be committed in completely different situations -- terrorist activities, domestic violence, sexual assault, riots, etc. -- but it is \textit{not} implied that a case on `terrorist activity' is similar to another on `domestic violence' even if both cases cite IPC Section 302.
For all the three statutes shown in Table~\ref{tab:icf-eg}, we find that the types of documents citing them vary widely.
%We understand that the problem of over-estimation of similarity is due to the fact that citations to all nodes, including the very generic nodes, are given equal importance in Hier-SPCNet-m2v. 

We understand that it is {\it not} desirable that citations to all nodes, including the very generic nodes, are given equal importance (which is what is done in Hier-SPCNet-m2v). 
So, we need a way to distinguish the discriminatory power of each node, when that node is cited by another node.

\begin{table}[tb]
\caption{Top 3 highly cited statutes in the Indian judiciary and some types of cases that have cited these statutes. The case type ($3^{rd}$ column) is not explicitly mentioned in the case documents; has been determined manually by law experts for these examples.}
\label{tab:icf-eg}
\centering
\small
\begin{tabular}{|c|c|l|}
\hline
\textbf{Statute}                                                                   & \textbf{Title of the Statute}                                                                         & \multicolumn{1}{c|}{\textbf{Types of cases that cite the Statute}}                                                 \\ \hline
\begin{tabular}[c]{@{}c@{}}Article 226 Constitution \\ of India, 1950\end{tabular} & \begin{tabular}[c]{@{}c@{}}Power of High Courts to \\ issue certain writs\end{tabular} & \begin{tabular}[c]{@{}l@{}}Land \& Property; Sales \& Indirect Tax; \\ Cancellation of Trade License;\end{tabular} \\ \hline
\begin{tabular}[c]{@{}c@{}}Section 302 Indian \\ Penal Code, 1860\end{tabular}     & Punishment for Murder                                                                  & \begin{tabular}[c]{@{}l@{}}Terrorist activities; Domestic violence;\\ Rape/Sexual Assault\end{tabular}                 \\ \hline
\begin{tabular}[c]{@{}c@{}}Article 14 Constitution \\ of India, 1950\end{tabular}  & Equality before Law                                                                    & \begin{tabular}[c]{@{}l@{}}Admission to colleges; Labour Wages;\\ Recruitment; Electoral Votes\end{tabular}             \\ \hline
\end{tabular}
\vspace{-5mm}
\end{table}

\vspace{1mm}
\noindent \textbf{Improving legal case similarity estimation:}
To address the limitation described above, we propose two modifications to Hier-SPCNet-m2v.

\noindent \textbf{(i) Inverse Citation Frequency (ICF):} 
We first propose to down-weight the frequently-cited statutes and document nodes. For a node $s$, we define $icf(s)$ as :
\begin{equation}
icf(s) = \log_{10} \left[\frac{N}{1+cf(s)}\right]
\end{equation}
where $cf(s)$ is the {\it citation frequency} of $s$, i.e., the number of times node $s$ is cited in the citation network (Hier-SPCNet), the number $1$ acts as a smoothing factor (to handle the case when $cf(s) = 0$), and $N$ is the total number of nodes that have outgoing citation links in Hier-SPCNet. 
The idea of ICF is analogous to that of Inverse Document Frequency in Information Retrieval.
While higher in-degree of a node implies higher importance in the citation network, higher in-degree has a reverse effect on the discriminatory power of a node.\footnote{A better indicator of the discriminatory power of a statute/case $s$ would have been the distribution of legal domains from which other cases have cited $s$. However, there is no easy way of knowing the legal domain of a case in our dataset.}

\noindent
\textbf{(ii) Hier-SPCNet-ICF-m2v (ICF-weighted-metapath2vec in Hier-SPCNet):}
Next, we modify the random walks of metapath2vec over Hier-SPCNet such that the nodes having low ICF contribute less to the final similarity score, thus attenuating the problem of over-estimation of similarity. 
Following Definition 4.1, consider the heterogeneous Hier-SPCNet graph as $G = (V,E,T)$, where $V$ is the set of nodes, $E$ is the set of edges, and $T$ denotes the different node types (section, topic, chapter, part, act, document). Let $\phi(v)$ denotes the node type $t\in T$ of $v$ and, let a metapath be represented by the schema $P = V_1 \overset{R_1}{\rightarrow} V_2 \overset{R_2}{\rightarrow} \cdots V_t \overset{R_t}{\rightarrow} V_{t+1} \cdots \overset{R_{l-1}}{\rightarrow} V_{l}$ (e.g., $doc \rightarrow sec \rightarrow act \rightarrow sec \rightarrow doc$). 
At time step $i$, let the random walker be at $v_t^i$ (vertex $v$ of type $t$). 
In the original metapath2vec formulation~\cite{dong2017metapath2vec}, the transition probability of the walker to another vertex $v^{i+1}$ (provided $v^{i+1}$ is of type $t+1$ ,i.e., $\phi (v^{i+1}) = t+1$) as defined by the metapath $P$, is a \textit{uniform distribution} among all the neighbours of $v_t^i$ that are of type $t+1$, denoted by $N_{t+1}(v_t^i)$.

In our proposed variant \textit{ICF-m2v}, every node $s$ in Hier-SPCNet is weighted equal to $icf(s)$. 
The \textit{transition probability of the walker from $v_t^i$ depends on the ICF weights of the neighbours of $v_t^i$}.
Suppose the random walker is at $v_t^i$ (vertex $v^i$ of type $t$). Under the metapath schema $P$, the walker at $v_t^i$ has to choose a vertex $v^{i+1}$ to transit given there is an edge $(v_t^i, v^{i+1}) \in E$ and  $\phi (v^{i+1}) = t+1$. 
If there are multiple nodes of type $t+1$ in the neighbourhood of $v_t^i$ and hence the walker has to choose among multiple $v^{i+1}$, then the ICF property of $v^{i+1}$ is considered. 
For example, if there are 2 neighbourhood nodes $v_m$ and $v_n$ of $v_t^i$, and $icf(v_m) > icf(v_n)$, then the probability of transiting to node $v_m$ is higher than that of transiting to $v_n$. Thus, the modified transition probability is:
\vspace{-2mm}
\begin{equation}
    \begin{split}
        p(v^{i+1}| v_t^i, P) & = \frac{icf(v^{i+1})}{\sum_{n=1}^{|N_{t+1}(v_t^i)|} icf(n)}, \; (v^{i+1},\; v_t^i) \in E, \phi(v^{i+1}) = t+1 \\
&=0,  (v^{i+1},\; v_t^i) \in E, \phi(v^{i+1}) \neq t+1 \\
&=0, (v^{i+1},\; v_t^i) \notin E
    \end{split}
\label{eqn:icf_m2v}
\end{equation}
Eqn.~\ref{eqn:icf_m2v} indicates that the transition probability from $v_t^i$ to $v^{i+1}$ following the metapath schema $P$ is directly proportional to $icf(v^{i+1})$, normalized by the $icf$ values of all the neighbourhood nodes $n$ (i.e., $icf(n)$) of $v_t^i$.
We use this transition probability, modified using the \textit{icf} values of the nodes, on the 14 metapaths of Hier-SPCNet-m2v for document similarity. We used the same settings of metapath2vec as Hier-SPCNet-m2v and obtain 200 dimensional embeddings of the nodes. The subsequent section explains how incorporating ICF improves upon the original metapath2vec based method.

\subsection{Performance analyses : Hier-SPCNet-ICF-m2v and Hier-SPCNet-m2v}
\label{ss:results-hspcnet-m2v-icf-m2v}
\input{sections/results-network-2}

%% file: sections/results-network-1.tex
\begin{table}[tb]
\centering
\caption{Results of the different similarity methods over PCNet and Hier-SPCNet on the Validation Set. Best results are in \textbf{bold}; second best results are \underline{underlined}. * indicates statistically significant improvement in Hier-SPCNet when compared to PCNet ($p<0.05$ as measured by paired Students T-Test for Correlation and MSE, Permutation test for FScore).}
\label{tab:pcnet-vs-hspcnet}
\begin{tabular}{|c|c|c|c|c|c|c|}
\hline
\multirow{2}{*}{\textbf{Method}} & \multicolumn{3}{c|}{\textbf{PCNet}} & \multicolumn{3}{c|}{\textbf{Hier-SPCNet}} \\ \cline{2-7} 
 & \textbf{Correlation} & \textbf{MSE} & \textbf{FScore} & \textbf{Correlation} & \textbf{MSE} & \textbf{FScore} \\ \hline
\begin{tabular}[c]{@{}c@{}}Bibliographic\\ Coupling\end{tabular} & 0.306 & 0.3576 & 0.355 & 0.573* & 0.3285* &  0.360\\ \hline
Co-citation & 0.226 & 0.2903 & 0.337 & 0.226 & 0.2903 & 0.337 \\ \hline
Dispersion & 0.209 & 0.2723 & 0.337 & 0.239 & 0.2935 & 0.337  \\ \hline
Node2Vec & 0.466 & 0.0841 & 0.672 & \underline{0.558*} & \underline{0.0647} & \underline{0.692} \\ \hline
Metapath2Vec & 0.451 & 0.0818 & 0.696 & \textbf{0.668}* & \textbf{0.0515}* & \textbf{0.705} \\ \hline
\end{tabular}
\vspace{-5mm}
\end{table}

\begin{table}[tb]
\centering
\caption{Results of the different similarity methods over PCNet and Hier-SPCNet on the Test Set. Best results are in \textbf{bold}; second best results are \underline{underlined}. * indicates statistically significant improvement in Hier-SPCNet when compared to PCNet ($p<0.05$ as measured by paired Students T-Test for Correlation and MSE, Permutation test for FScore).}
\label{tab:pcnet-vs-hspcnet-test}
\begin{tabular}{|c|c|c|c|c|c|c|}
\hline
\multirow{2}{*}{\textbf{Method}} & \multicolumn{3}{c|}{\textbf{PCNet}} & \multicolumn{3}{c|}{\textbf{Hier-SPCNet}} \\ \cline{2-7} 
 & \textbf{Correlation} & \textbf{MSE} & \textbf{FScore} & \textbf{Correlation} & \textbf{MSE} & \textbf{FScore} \\ \hline
\begin{tabular}[c]{@{}c@{}}Bibliographic\\ Coupling\end{tabular} & 0.287 & 0.3821 & 0.317 & 0.505* & 0.3078* &  0.329\\ \hline
Co-citation & 0.205 & 0.3173 & 0.309 & 0.205 & 0.3173 & 0.205 \\ \hline
Dispersion & 0.188 & 0.2967 & 0.284 & 0.225 & 0.2845 & 0.316  \\ \hline
Node2Vec & 0.441 & 0.0815 & 0.604 & \underline{0.514*} & \underline{0.0745} & \underline{0.619} \\ \hline
Metapath2Vec & 0.454 & 0.0756 & 0.635 & \textbf{0.602}* & \textbf{0.0498}* & \textbf{0.644} \\ \hline
\end{tabular}
\vspace{-5mm}
\end{table}

We now compare the performance of the different network-based metrics over the two networks -- PCNet and proposed Hier-SPCNet. Table~\ref{tab:pcnet-vs-hspcnet} shows the results over the validation set, over 100 document pairs. Table~\ref{tab:pcnet-vs-hspcnet-test} shows the results over the test set, over 90 document pairs.
As stated earlier, all hyper-parameters are decided based on the performance of various methods on the validation set; the same hyper-parameter values are used when the methods are applied on the test set (no further tuning done on the test set).
As detailed in Section~\ref{sub:evaluation-metrics}, we report Correlation, MSE, and F-Score. 

As expected, the performances of almost all methods are slightly lower over the test set than over the validation set; however, the trends are similar over both datasets.
In both the datasets, all the methods show improved scores when applied on Hier-SPCNet than when applied on PCNet, except for the method `co-citation'. This is because, `co-citation' is calculated based on the number of common {\it in-citations}. Since, in-citations to a case document can only come from other \textit{case documents} and not statutes, the value is same for both PCNet and Hier-SPCNet.

In particular, the value of bibliographic coupling in Hier-SPCNet is significantly higher than PCNet. This shows that \textit{statutes and precedents} are important for correctly estimating the similarity between two case documents.
Also, the node embedding techniques Node2Vec and Metapath2Vec show substantial improvement over Hier-SPCNet than measures used in prior-work.

% Although Node2Vec considers the graph to be homogeneous, including the hierarchical structure of statutes over PCNet helps, since the leaf nodes, i.e., the \textit{section/article} nodes are structurally similar. 

Metapath2vec over Hier-SPCNet shows the best performance across all measures and over both the validation set and test set, with a correlation of $0.668$ with mean expert similarity score over the validation set and a correlation of $0.602$ over the test set. 
These results show that the metapaths are more efficient at capturing the similarity between case documents. The schemas are able to encode the legal knowledge inherent in the statutes. 

It can be noted that the metapaths and the Hier-SPCNet network were defined in consultation with the same Law experts (from the RGSOIPL Law school) who annotated the validation set. However, the document-pairs in the test set were annotated by a completely different set of Law experts, from a different Law school (WBNUJS). 
Hence, the results over the test set in particular indicates that the proposed method of using Metapath2vec over Hier-SPCNet can well capture the notion of legal document similarity as agreed upon by various groups of Law experts.

%% file: sections/results-network-2.tex
% \begin{table}[]
% \caption{Performance comparison of Hier-SPCNet-m2v and Hier-SPCNet-ICF-m2v. Improvements are not statistically significant at p<0.05 as measured by Paired Students' T-Test.}
% \label{tab:m2v-icfm2v}
% \begin{tabular}{|c|c|c|c|}
% \hline
% \textbf{Method} & \textbf{Correlation} & \textbf{MSE} & \textbf{FScore} \\ \hline
% Hier-SPCNet-m2v & 0.652 & 0.0580 & 0.711 \\ \hline
% Hier-SPCNet-ICF-m2v & 0.708 & 0.0507 & 0.752 \\ \hline
% \end{tabular}
% \end{table}

% \begin{table}[tb]
% \caption{Performance comparison of Hier-SPCNet-m2v and Hier-SPCNet-ICF-m2v. Improvements are not statistically significant at p<0.05 as measured by Paired Students' T-Test.}
% \label{tab:m2v-icfm2v}
% \vspace{-2mm}
% \begin{tabular}{|c|c|c|c|}
% \hline
% \textbf{Method} & \textbf{Correlation} & \textbf{MSE} & \textbf{FScore} \\ \hline
% Hier-SPCNet-m2v & 0.668 & 0.0515 & 0.705  \\ \hline
% Hier-SPCNet-ICF-m2v & \textbf{0.725} & \textbf{0.0427} & \textbf{0.779}  \\ \hline
% \end{tabular}
% \vspace{-4mm}
% \end{table}

\begin{table}[tb]
\centering
\caption{Performance comparison of Hier-SPCNet-m2v and Hier-SPCNet-ICF-m2v on the validation and test sets. Hier-SPCNet-ICF-m2v achieves better performance according to all metrics over both datasets. However, improvements are not statistically significant at p<0.05 as measured by Paired Students' T-Test.}
\label{tab:m2v-icfm2v}
\begin{tabular}{|c|ccc|ccc|}
\hline
\multirow{2}{*}{\textbf{Method}} & \multicolumn{3}{c|}{\textbf{Validation Set}}                                                & \multicolumn{3}{c|}{\textbf{Test Set}}                                                      \\ \cline{2-7} 
                                 & \multicolumn{1}{c|}{Correlation}    & \multicolumn{1}{c|}{MSE}             & FScore         & \multicolumn{1}{c|}{Correlation}    & \multicolumn{1}{c|}{MSE}             & FScore         \\ \hline
Hier-SPCNet-m2v                  & \multicolumn{1}{c|}{0.668}          & \multicolumn{1}{c|}{0.0515}          & 0.705          & \multicolumn{1}{c|}{0.602}          & \multicolumn{1}{c|}{0.0498}          & 0.644          \\ \hline
Hier-SPCNet-ICF-m2v              & \multicolumn{1}{c|}{\textbf{0.725}} & \multicolumn{1}{c|}{\textbf{0.0427}} & \textbf{0.779} & \multicolumn{1}{c|}{\textbf{0.650}} & \multicolumn{1}{c|}{\textbf{0.0402}} & \textbf{0.665} \\ \hline
\end{tabular}
\end{table}

We apply both m2v and ICF-m2v on Hier-SPCNet and determine the similarity scores of the document-pairs in the two datasets.
Table~\ref{tab:m2v-icfm2v} shows the performance of the two methods in terms of Pearson correlation, MSE and F-Score with respect to expert given scores. The first set of three columns are for the validation set, while the next set of columns are for the test set. 
We find that Hier-SPCNet-ICF-m2v which takes into account the discriminatory power of the nodes, performs better than Hier-SPCNet-m2v across all evaluation measures, over both the validation and the test datasets. 

Table~\ref{tab:doc-pair-eg} shows some document-pairs for which Hier-SPCNet-ICF-m2v estimates better similarity scores (i.e, scores that are closer to the expert-assigned similarity scores) than Hier-SPCNet-m2v.
For instance, consider the document pair 1995\_S\_317 \& 2011\_I\_16 in Table~\ref{tab:doc-pair-eg}. They are very different according to legal experts (expert similarity score is $0.03$). This is because even if both the documents are related to criminal offences and murder, one is about prevention of corruption (Section 13 Prevention of Corruption Act cited by 1995\_S\_317) and the other is about terrorist activities (Section 3 Terrorist and Disruptive Activities (Prevention) Act cited by 2011\_I\_16.)
We observe that, both the documents cite Section 302 of the Indian Penal Code (IPC) which talks about `Punishment for murder'. But this section is very commonly cited by any criminal case involving a murder. The ICF of IPC Section 302 is therefore relatively low. The other common sections cited by the two case documents also have lower ICFs and therefore Hier-SPCNet-ICF-m2v could attenuate the similarity to $0.332$, which is closer to the expert score. Since Hier-SPCNet-m2v considers all the sections with equal importance, it infers a higher similarity value of $0.494$. 
Similar arguments hold for the document-pair (1961\_34 \& 1987\_37) in Table~\ref{tab:doc-pair-eg}, which cite articles related to Fundamental Rights. These articles are also very generic and lack the ability to discriminate between document pairs.

\begin{table}[tb]
\centering
\caption{Examples of some document pairs whose similarity was over-estimated or under-estimated by Hier-SPCNet-m2v but have been corrected by Hier-SPCNet-ICF-m2v.}
\label{tab:doc-pair-eg}
\scalebox{0.5}{
\begin{tabular}{|c|l|l|c|l|l|}
\hline
\textbf{\begin{tabular}[c]{@{}l@{}}Docs in\\ the Pair\end{tabular}} & \multicolumn{1}{c|}{\textbf{Statutes Cited}} & \textbf{\begin{tabular}[c]{@{}l@{}}Common\\ Precedent\end{tabular}} & \multicolumn{1}{c|}{\textbf{\begin{tabular}[c]{@{}c@{}}Some Relevant\\ Metapaths\end{tabular}}}& \multicolumn{1}{c|}{\textbf{\begin{tabular}[c]{@{}c@{}}Demonstration of how the metapaths capture\\ the similarity between the documents\end{tabular}}} & \multicolumn{1}{c|}{\textbf{\begin{tabular}[c]{@{}c@{}}ICF values of the \\ Statutes \& Precedent \end{tabular}}} \\ \hline

\multicolumn{6}{|c|}{\textit{Expert Similarity Score = 0.03 ; Network Similarity inferred by (i) Hier-SPCNet-m2v = 0.494 (ii) Hier-SPCNet-ICF-m2v = 0.332}} \\ \hline
\multirow{2}{*}{1995\_S\_317} & \multirow{2}{*}{\begin{tabular}[c]{@{}l@{}}Indian Penal Code 1860 (IPC)\_147, IPC\_148,\\ IPC\_149, IPC\_307, IPC\_302, IPC\_323, IPC\_324,\\ Constitution\_136, Constitution\_142, Terrorist and \\ Disruptive Activities Prevention Act 1987 (TADA)\_3\end{tabular}} & \multicolumn{1}{c|}{\multirow{3}{*}{--}} & doc-sec-doc & \begin{tabular}[c]{@{}l@{}}(i) 1995\_S\_317 -- IPC\_302 -- 2011\_I\_16\\ (ii) 1995\_S\_317 -- Constitution\_142 -- 2011\_I\_16\end{tabular} & \multirow{3}{*}{\begin{tabular}[c]{@{}l@{}}Constitution\_145 : 6.06\\ IPC\_324 : 4.27, Constitution\_142 : 4.21\\ IPC\_147 : 4.18 , IPC\_323 : 4.09\\ IPC\_148 : 3.83 , IPC\_307 : 3.73\\ IPC\_149 : 3.58, Constitution\_136 : 2.77\\ IPC\_302 : 2.43\end{tabular}} \\ \cline{4-5}
 &  & \multicolumn{1}{c|}{} & doc-sec-chapter-sec-doc & \begin{tabular}[c]{@{}l@{}}1995\_S\_317 -- Constitution\_142 --\\ Constitution\_Chapter 4 -- \\Constitution\_145 -- 2011\_I\_16\end{tabular} &  \\ \cline{1-2} \cline{4-5}
2011\_I\_16 & \begin{tabular}[c]{@{}l@{}}IPC\_302 ; Constitution\_142, Constitution\_145,\\ Prevention of Corruption Act 1988 (PCA)\_7,\\ PCA\_13\end{tabular} & \multicolumn{1}{c|}{} & \begin{tabular}[c]{@{}l@{}}doc-sec-topic-chapter-topic-\\ sec-doc\end{tabular} & \begin{tabular}[c]{@{}l@{}}1995\_S\_317 -- IPC\_324 -- IPC\_Topic 4 --\\ IPC\_Chapter 16 -- IPC\_Topic 2 -- IPC\_302 -- 2011\_I\_16\end{tabular} &  \\ \hline
\multicolumn{6}{|c|}{\textit{Expert Similarity Score = 0.07; Network Similarity inferred by (i) Hier-SPCNet-m2v = 0.372 (ii) Hier-SPCNet-ICF-m2v = 0.226}} \\ \hline

1961\_34 & Constitution\_14, Constitution\_20 , Constitution\_32 & \multicolumn{1}{c|}{\multirow{3}{*}{--}} & doc-sec-topic-sec-doc & \begin{tabular}[c]{@{}l@{}}1961\_34 -- Constitution\_20 -- Constitution\_topic2 --\\ Constitution\_21 -- 1987\_37\end{tabular} & \multirow{3}{*}{\begin{tabular}[c]{@{}l@{}}Constitution\_22 : 4.62\\ Constitution\_21 : 3.70\\ Constitution\_32 : 2.89\\ Constitution\_14 : 2.64\end{tabular}} \\ \cline{1-2} \cline{4-5}
\multirow{2}{*}{1987\_37} & \multirow{4}{*}{\begin{tabular}[c]{@{}l@{}}Constitution\_21, Constitution\_22, Gujarat Prevention of\\Antisocial Activities Act 1985 (GPSA)\_10,GPSA\_15,\\National Security Act 1980\_11, IPC\_300, IPC\_302,\\ IPC\_303, Code of Criminal Procedure 1973 (CrPC)\_235,\\ CrPC\_354, CrPC\_397\end{tabular}} & \multicolumn{1}{c|}{} & \multirow{4}{*}{\begin{tabular}[c]{@{}l@{}}doc-sec-topic-part-\\ topic-sec-doc\end{tabular}} & \multirow{4}{*}{\begin{tabular}[c]{@{}l@{}}(i) 1961\_34 -- Constitution\_14 -- Constitution\_Topic1\\  -- Constitution\_Part III  -- Constitution\_Topic 2--\\ Constitution\_21 -- 1987\_37 \\ (ii) 1961\_34 -- Constitution\_32 -- Constitution\_Topic 7\\ -- Constitution\_Part III -- Constitution\_Topic 2--\\ Constitution\_22 -- 1987\_37\end{tabular}} &  \\ 
 &  &  &  &  &  \\ %\cline{4-5}
  &  &  &  &  &\\
  &  &  &  &  &\\
  &  &  &  &  &\\

 &  & \multicolumn{1}{c|}{} &  &  &  \\ \hline
 \multicolumn{6}{|c|}{\textit{Expert Similarity Score = 0.87 ; Network Similarity inferred by (i) Hier-SPCNet-m2v = 0.579 (ii) Hier-SPCNet-ICF-m2v = 0.684}} \\ \hline
\multirow{4}{*}{1985\_113} & \multirow{4}{*}{\begin{tabular}[c]{@{}l@{}}National Security Act 1980 (NSA)\_3, NSA\_10,\\ NSA\_11, NSA\_12,  Constitution\_226,\\ Constitution\_21, Conservation of Foreign\\ Exchange and Prevention of Smuggling\\ Activities Act (COFEPOSA)\_8\end{tabular}} & \multirow{5}{*}{1969\_324} & doc-sec-doc & 1991\_12 -- COFEPOSA\_8 -- 1985\_113 & \multirow{5}{*}{\begin{tabular}[c]{@{}l@{}}COFEPOSA\_8 : 9.55\\ COFEPOSA\_3 : 5.75\\ Constitution\_22 : 4.62\\ Constitution\_21 : 3.70\\ 1969\_324 : 5.00\end{tabular}} \\ \cline{4-5}
 &  &  & \multirow{2}{*}{doc-sec-topic-sec-doc} & \multirow{2}{*}{\begin{tabular}[c]{@{}l@{}}1991\_12 -- Constitution\_22 --Constitution\_topic2 --\\ Constitution\_21 -- 1985\_113\end{tabular}} &  \\
 &  &  &  &  &  \\ \cline{4-5}
 &  &  & doc-sec-act-sec-doc & \begin{tabular}[c]{@{}l@{}}1991\_12 -- COFEPOSA\_3 -- COFEPOSA --\\ COFEPOSA\_8 -- 1985\_113\end{tabular} &  \\ \cline{1-2} \cline{4-5}
1991\_12 & \begin{tabular}[c]{@{}l@{}}Preventive Detention Act 1950\_3,\\ Constitution\_22,  General Clauses Act 1987\_21,\\ COFEPOSA\_3, COFEPOSA\_8\end{tabular} &  & doc-doc-doc & 1991\_12 -- 1969\_324 -- 1985\_113 &  \\ \hline
\end{tabular}
}
\end{table}

Hier-SPCNet-ICF-m2v not only decreases the similarity scores, but also increases the similarity scores in scenarios when the common statues cited by two documents have higher ICF values. Such scenarios imply that the two documents cite similar statutes, which are cited in specific legal issues only and are not very generic. 
Therefore the documents are actually discussing similar legal issues and can be inferred to be similar with a higher confidence. An example is the pair 1985\_113 and 1991\_12 (third example in Table~\ref{tab:doc-pair-eg}) where 
Section 8 of the COFEPOSA Act (cited by both the documents) has a very high ICF value. 
Here Hier-SPCNet-ICF-m2v is able to infer a better similarity score of $0.684$ (closer to the expert-assigned similarity score of $0.87$) than Hier-SPCNet-m2v (which infers $0.579$).

%The expert-judged similarity score between 1985\_113 and 1991\_12 is $0.87$ which is closer to the similarity score inferred by Hier-SPCNet-ICF-m2v ($0.684$) than to what is inferred by Hier-SPCNet-m2v ($0.579$).

%% file: sections/text-mtds.tex
Signals of similarity also comes from the text/content of the documents. In this section, we describe methods for estimating text-based or content-based legal document similarity. 
There are two broad approaches for text-based similarity -- supervised and unsupervised. 

%\vspace{2mm}
%\noindent \textbf{Challenges in applying supervised text-based similarity methods}: 
\subsection{Challenges in applying supervised text-based similarity methods}

As mentioned in Section~\ref{ss:relatedwork-txt-mtds}, there exists generic document similarity methods that can be potentially applied to legal documents. However, there are several differences of the documents on which those methods have been applied (e.g. news articles, email collections), with legal case documents -- 
(i)~Legal case documents are very long as compared to news articles, emails, or even Wikipedia articles on which the above-stated methods have been applied.
For instance, the CNN/DM news articles have 656 words on average~\cite{cnndm}, emails from the the Avocado Research Email Collection have 112 words on average~\cite{avocado}), and Wikipedia articles have 624 words on average\footnote{\url{https://en.wikipedia.org/wiki/Wikipedia:Size_of_Wikipedia}}). Whereas, from the Indian Supreme Court jurisdiction, Civil cases are of 2,995 words and 140 sentences on average; Criminal cases are of 3,000 words and 145 sentences on average.
(ii)~The legal documents are multi-faceted, but these facets are not explicitly mentioned in the documents (at least in the Indian judiciary), unlike scientific articles; specifically, the documents in the Indian judiciary are unstructured text without any section headings. 
(iii)~It is very expensive to construct large-scale training data for supervised methods in the legal domain, since developing gold standard annotations require the involvement of law experts. Since legal documents are long, reading and understanding the documents to rate their similarity is time consuming. Additionally, since the task of document similarity is subjective in nature, therefore it requires similarity estimates from multiple experts, which makes it prohibitively expensive.

Nevertheless, we still explored supervised methods for the task. To this end, we attempted to curate training data pairs using distant-supervision techniques. We consulted our law experts for how to generate training data. 
The law experts opined that one document citing another can be taken as a confirmatory signal that the two documents are similar. Hence, for generating positive training samples, we randomly sampled two documents $d_i$ and $d_j$ from the collection. If there is a citation link between $d_i$ and $d_j$, we considered $(d_i, d_j)$ to be similar and hence a positive sample. 
For obtaining negative training samples (dissimilar document-pairs), we randomly sampled two documents $d_i$ and $d_j$. Then we took one of the three approaches: 
(i)~if there is no citation link between $(d_i, d_j)$, we consider it to be dissimilar. 
(ii)~if the Jaccard similarity between the outgoing citations from the two nodes is less than $0.2$, we considered the pair to be a negative example. 
(iii)~if the Jaccard similarity between the outgoing citations of the two nodes is less than $0.2$ and their textual similarity is less than $0.4$, we considered the pair to be a negative example. 
The thresholds 0.2 and 0.4 were decided based on the distribution of the Jaccard similarity and textual similarity values of a large number of document-pairs. We observed that a large majority of document-pairs have similarity values lesser than these thresholds, and only a relatively fraction of document-pairs have similarity values above these thresholds. Since most document-pairs are likely to be dissimilar, it seemed safe to consider the document-pairs lying below these thresholds to be dissimilar.

We then trained supervised methods for long document similarity estimation, such as SMASH-RNN~\cite{smash_rnn} and ConceptGraph~\cite{article_matching}, on a large synthetically generated training set containing similar and dissimilar document-pairs curated using the methods described above. 
However, we observed that the methods do \textit{not} perform well on the validation/test sets. Hence, we do not report the results of these supervised methods.

%\vspace{2mm}
%\noindent \textbf{Unsupervised method for text-based legal document similarity}: 
\subsection{Unsupervised methods for text-based legal document similarity}

We applied several unsupervised, text-based methods for legal document similarity. We describe these methods below.

\vspace{1mm}
\noindent $\bullet$ \textbf{Doc2Vec}: Doc2Vec~\cite{doc2vec}, is an unsupervised method for measuring the similarity between two documents/paragraphs. 
Recent studies have found Doc2Vec to be very effective for estimating the similarity between legal documents~\cite{mandal2017measuring,mandal-ailaw}
and also effective in the medical domain~\cite{d2v-btr-1}.
%Prior work has shown Doc2Vec to perform well in the legal~\cite{mandal-ailaw,d2v-btr-2} and medical~\cite{d2v-btr-1} domain than supervised methods. 
We therefore use Doc2vec for measuring the similarity between two legal documents, following the approach used in~\cite{mandal-ailaw}.
We perform basic preprocessing steps as listed in~\cite{mandal-ailaw} such as stopword removal, stemming, converting all words to lowercase, etc.
Then we train Doc2Vec on $53,068$ Indian Supreme Court case documents, excluding the documents in the validation and test sets.
We use the Doc2Vec implementation from the open-source package Gensim\footnote{\url{https://radimrehurek.com/gensim/models/doc2vec.html}}.
The embedding dimension is taken as $200$ (which gave the best results over the validation set) and other parameters are set to default. 
Finally, to find the similarity between the document pair $(d_1,d_2)$, we compute cosine similarity between the inferred Doc2vec embeddings of the documents $d_1$ and $d_2$. 
In the rest of the paper, we refer to this method as \textbf{Doc2Vec}.

\vspace{1mm}
\noindent $\bullet$ \textbf{Pretrained Transformer models}: Recently, pretrained transformer models like Bert~\cite{devlin-bert} have gained a lot of popularity owing to their transfer learning capability. Given a small amount of task-specific training data, these models have shown significant improvements in various NLP tasks.

We apply four transformer models -- Bert~\cite{devlin-bert}, LegalBert~\cite{chalkidis2020legal}, BERT-PLI~\cite{bert-pli} and RoBERTa~\cite{roberta} -- for the task of legal document similarity. 
In the absence of training data to fine-tune these models, we use the pre-trained models in an unsupervised manner. For a particular document-pair ($d_1$, $d_2$), we obtain their embeddings ($e_1$, $e_2$) using the pre-trained transformer models as follows.

As described in earlier sections, legal documents are inherently lengthy. It is known that most transformer models have a length restriction of the number of input tokens (e.g. 512 tokens for Bert), which is far less than the length of a legal document. In a prior work~\cite{mandal-ailaw} this limitation was handled as follows -- every document was broken down into chunks, where a chunk is defined as a set of consecutive sentences. In order to maintain the sequence information across consecutive chunks, the last few sentences from the previous chunk are added to the beginning of the next chunk. Each chunk is then fed to the transformer model and a chunk-embedding is obtained. The chunk-embeddings of all the chunks in a document are then averaged to get the document-embedding (for the whole document). We adopt the same procedure in this work. We construct chunks of size 5 sentences, including  two sentences from the previous chunk and the next three sentences. 
The document-embedding is obtained by averaging the chunk-embeddings obtained from one of the pretrained transformer models.
Once we obtain the document-embeddings ($e_1$, $e_2$) for a given document-pair,
we compute cosine similarity between the embeddings ($e_1$, $e_2$) to get the final similarity score between the said document-pair.

We use the Huggingface implementations of the models BERT (bert-base-uncased), LegalBert (legal-bert-base-uncased) and RoBERTa (roberta-base). For BERT-PLI, we could not find the trained model available from the original work~\cite{bert-pli}. We found a reproducibility paper~\cite{bert-pli-repro} that replicates the model architecture and makes the trained model (trained over the same dataset as in the original work) available.\footnote{\url{https://zenodo.org/record/4088010\#.YqsurS8RoVU}; in particular, we use the model bert-pli-reproduction/lawbert/pytorch\_model.bin in our experiments.} We used this trained model for BERT-PLI for our experiments.

\vspace{2mm}
\noindent The results of the text-based similarity methods are shown in Table~\ref{tab:text-results}. We find that Doc2Vec performs the best in all the evaluation metrics, over both the validation and test sets. Roberta performs better than the other transformer-based models in terms of correlation, while LegalBERT performs better in terms of MSE. Among the models trained on legal corpus (Legalbert and BERT-PLI), we find LegalBert to be performing better.
In general, we observed that all the four transformer-based models infer too high similarity scores for almost all the document-pairs, thus leading to poor correlation and higher MSE values.

It can be noted that the correlation score reported for Bert in~\cite{mandal-ailaw} is similar to what is observed in our experiments. Additionally, Doc2vec was observed to be better than a Bert-based chunking approach for estimating legal document similarity also in~\cite{mandal-ailaw}.

%Doc2vec performs marginally better than the best network-based method discussed in earlier sections.

% \begin{table}[t]
% \centering
% \caption{Results of the text-based method (Doc2Vec) and best performing network-based method (Hier-SPCNet-ICF-m2v; values repeated from Table~\ref{tab:m2v-icfm2v}) for computing  legal case document similarity.}
% \label{tab:text-results}
% \begin{tabular}{|c|c|c|c|c|}
% \hline
% \textbf{Type} & \textbf{Method} & \textbf{Correlation} & \textbf{MSE} & \textbf{F-score} \\ \hline
 
%  Text-based & Doc2Vec & 0.777 & 0.0381 & 0.774  \\ \hline
%  Network-based & Hier-SPCNet-ICF-m2v &  0.725 & 0.0427 & 0.779  \\ \hline
 
% \end{tabular}
% \vspace{-6mm}
% \end{table}

% \usepackage{multirow}
% \usepackage[normalem]{ulem}
\begin{table}[t]
\centering
%\caption{\new{Results of the text-based methods and best performing network-based method (Hier-SPCNet-ICF-m2v; values repeated from Table~\ref{tab:m2v-icfm2v}) on the Validation and Test sets respectively.}}
\caption{Results of unsupervised (pretrained) text-based methods on the validation set and test set. The best values are in boldface and the second-best values are underlined.}
\label{tab:text-results}
\begin{tabular}{|c|c|ccc|ccc|}
\hline
\multirow{2}{*}{\textbf{Type}} & \multirow{2}{*}{\textbf{Method}} & \multicolumn{3}{c|}{\textbf{Validation Set}} & \multicolumn{3}{c|}{\textbf{Test Set}} \\ \cline{3-8} 
 &  & \multicolumn{1}{c|}{Correlation} & \multicolumn{1}{c|}{MSE} & FScore & \multicolumn{1}{c|}{Correlation} & \multicolumn{1}{c|}{MSE} & FScore \\ \hline
\multirow{5}{*}{Text-based} & Doc2Vec & \multicolumn{1}{c|}{\textbf{0.765}} & \multicolumn{1}{c|}{\textbf{0.039}} & \textbf{0.768} & \multicolumn{1}{c|}{\textbf{0.701}} & \multicolumn{1}{c|}{\textbf{0.0356}} & \textbf{0.682} \\ \cline{2-8} 
 & BERT & \multicolumn{1}{c|}{0.207} & \multicolumn{1}{c|}{0.367} & 0.309 & \multicolumn{1}{c|}{0.198} & \multicolumn{1}{c|}{0.372} & 0.314 \\ \cline{2-8} 
 & LegalBert & \multicolumn{1}{c|}{0.290} & \multicolumn{1}{c|}{\underline{0.351}} & 0.315 & \multicolumn{1}{c|}{0.301} & \multicolumn{1}{c|}{\underline{0.332}} & \underline{0.320} \\ \cline{2-8} 
 & BERT-PLI & \multicolumn{1}{c|}{0.265} & \multicolumn{1}{c|}{0.367} & \multicolumn{1}{c|}{0.310} & \multicolumn{1}{c|}{0.278} & \multicolumn{1}{c|}{0.341} & \multicolumn{1}{c|}{0.312} \\ \cline{2-8} 
 & RoBERTa & \multicolumn{1}{c|}{\underline{0.351}} & \multicolumn{1}{c|}{0.372} & \underline{0.328} & \multicolumn{1}{c|}{\underline{0.334}} & \multicolumn{1}{c|}{0.398} & 0.305 \\ \hline
%\begin{tabular}[c]{@{}c@{}}Network-\\ based\end{tabular} & \begin{tabular}[c]{@{}c@{}}Hier-SPCNet-\\ ICF-m2v\end{tabular} & \multicolumn{1}{c|}{0.725} & \multicolumn{1}{c|}{0.0427} & 0.779 & \multicolumn{1}{c|}{0.65} & \multicolumn{1}{c|}{0.0402} & 0.665 \\ \hline
\end{tabular}
\end{table}

Although we have experimented with supervised methods for measuring document similarity (as described above), we found better performance with Doc2Vec. It is possibly because the model was provided with a collection of $53,068$ case documents, albeit unannotated, whereby it could learn the document vectors well. 
The supervised methods, on the other hand, were provided document-pairs for training that were generated through distant supervision techniques based on statistical methods, in absence of expert-annotated data. While this setup is not an ideal way of training supervised models, it is practically impossible to annotate thousands of document-pairs by domain experts in this domain.
The difference in performances brings forth an inherent disadvantage of supervised methods requiring large training data, since it is very expensive to obtain such large training data in expert-driven domains such as the legal domain.

\if 0 
\new{It can be noted that the models LegalBert and BERT-PLI are pretrained on legal text, but from jurisdictions other than India. 
Having transformer models pretrained specifically over a large volume of Indian legal data can possibly further improve the results on this task. 
However, such pretraining would require a huge amount of computing resources. This is left as a potential future work.}
\fi

%% file: sections/text+network-mtds.tex
In the previous sections we saw that there are two sources for signals on legal document similarity -- Network-based (Section~\ref{sec:nw-mtds}) and Text-based (Section~\ref{sec:text-mtds}). Our experiments showed Hier-SPCNet-ICF-m2v to be the best performing network-based method for capturing document similarity (see Table~\ref{tab:m2v-icfm2v}). Among the text-based methods, we consider Doc2Vec (see Table~\ref{tab:text-results}).
Now we attempt to combine the above information to get an aggregated view of document similarity. To the best of our knowledge, this is the first work that aims to combine network and text-based signals for capturing legal document similarity. 
We describe a variety of methods for this task below.

\subsection{Combining network-based and text-based similarity values}

We start with the simplest method -- inferring similarity values \textit{independently} by the text-based method and the network-based method, and then combining the two similarity values.

\noindent \textbf{~(I) Value Combination}: In this technique, for a document pair $(d_1,d_2)$ we calculate their textual similarity $text\_sim \! = \! cosine\_similarity(t_1,t_2)$ and network similarity  $nw\_sim \! = \! cosine\_similarity(n_1,n_2)$ separately. 
We compute the final similarity value in one of two ways:\\
\underline{(i) Value-Average}: $final\_sim \! = \! (text\_sim + nw\_sim)/2$ \\
\underline{(ii) Value-Max}: $final\_sim \! = \! max(text\_sim, nw\_sim)$. 

%\new{The above technique of combining textual and network information seem to be too straight-forward and simple. Since the similarity values are derived from the embeddngs, we take a step backward and attempt to explore methods that can take as input the embeddings themselves and combine them jointly to derive a single similarity score. With this hypothesis, we explore several embedding combination techniques both non-neural network Embedding combination (II) and neural network based Neural Network based combination (IV). Since we want to combine textual embeddings and embeddings coming from a \textit{network}, we investigate as to why not give the network as input coupled with the textual information embedded as node features? In this attempt, we use Graph-based Combination techniques (III). We describe each of these methods in detail subsequently.}

Note that, in the above approach, the text-based similarity and the network-based similarity are computed independently (before being combined).
Another alternative approach can be to do the combination first, i.e., to compute a single joint/composite embedding (representation) for a particular document $d$ from its text embedding and its network embedding. Once we get the joint/composite embeddings for both $d_1$ and $d_2$, we can measure the similarity between the two joint embeddings. The next section details this approach.

\begin{figure}[t]
	\centering
	\includegraphics[width=\columnwidth,height=6cm]{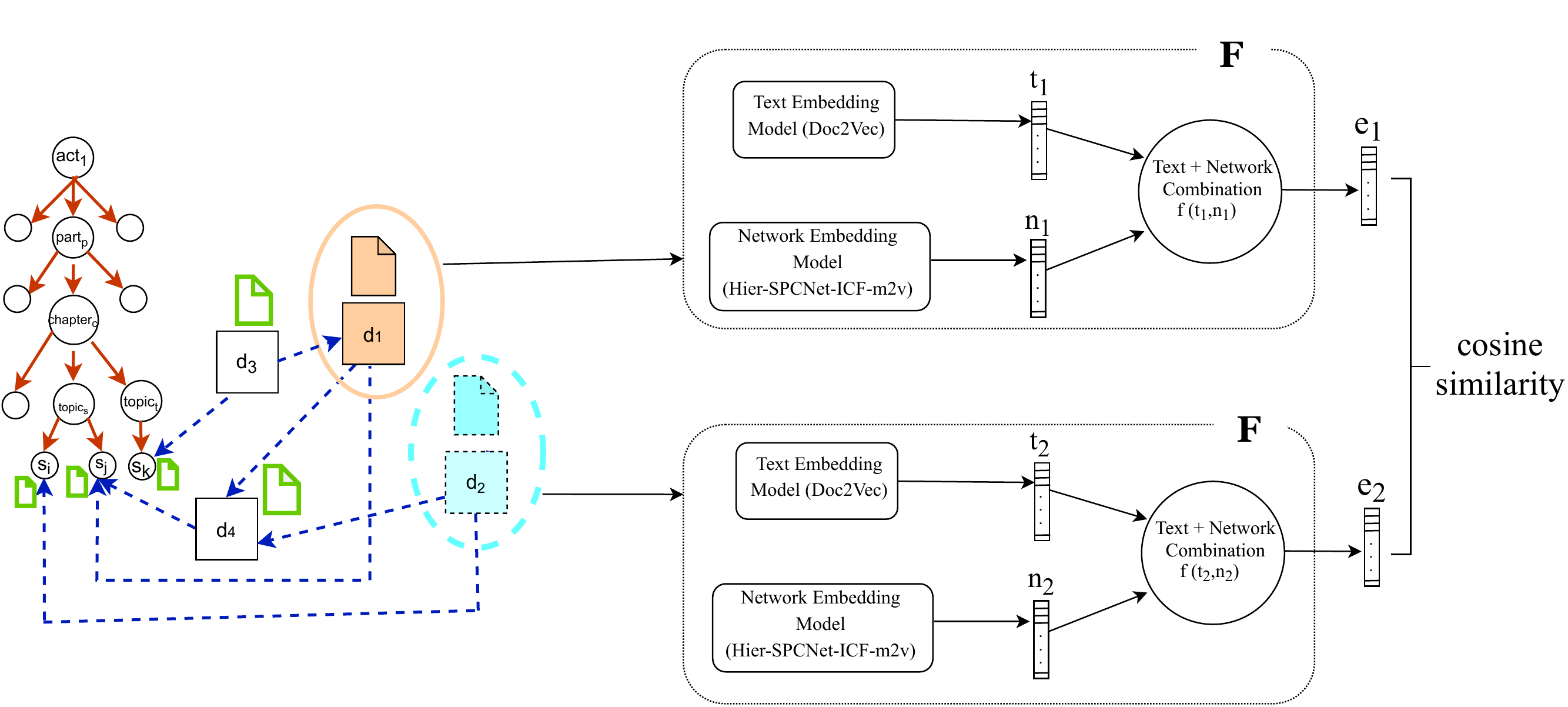}
	\caption{Workflow for computing similarity between two legal documents, by combining network and text embeddings. The input consists of two documents $d_1$ (shown in orange) and $d_2$ (shown in blue) whose similarities are to be computed. Both $d_1$ and $d_2$ have textual information (denoted corner-folded note symbols) and is present as a node in Hier-SPCNet. Both $d_1$ and $d_2$ are given as input to the function/module \textbf{F} that constructs a joint embedding $e_i$ by combining the textual embedding $t_i$ and network embedding $n_i$ for document $d_i$, for $i = 1,2$. The similarity between the documents is the cosine similarity between $e_1$ and $e_2$.}
	\label{fig:workflow}
	\vspace{-8mm}
\end{figure}

\subsection{Combining network and text embeddings}
\label{ss:txt+nw-mtds-desc}

Here we assume that the input is a document pair $(d_1,d_2)$ whose similarity is to be computed. Both the documents are nodes in Hier-SPCNet, and have associated text.
Each document is first passed through a function \textbf{F}, to get a joint embedding $e_1$ for $d_1$ and $e_2$ for $d_2$. The final similarity is computed as $cosine\_similarity (e_1,e_2)$.
The general workflow of the methods discussed in this section is illustrated in Figure~\ref{fig:workflow}.

The function \textbf{F} takes as input a document $d$, which has a textual content (indicated by a corner-folded note in Figure~\ref{fig:workflow}).
%\footnote{The leaf nodes $s_i$, $s_j$ and $s_k$ in the Statute hierarchy of Hier-SPCNet also has a textual content. We shall see how this information is used later in Section~\ref{ss:txt+nw-mtds-desc}(iii).}
The document is also a node in Hier-SPCNet, from which it is possible to understand its citation network structure. The textual information is sent to a Text Embedding model, which in this section is the Doc2Vec model as it was shown to have the best performance (refer Table~\ref{tab:text-results}), to get a text-based representation $t$ of the document. 
Our best performing network similarity model Hier-SPCNet-ICF-m2v (refer Table~\ref{tab:m2v-icfm2v}) infers the network embedding $n$ of the document $d$. 
The textual embedding $t$ and the network embedding $n$ of the document are aggregated through a function \textbf{$f(t,n)$} that outputs a joint embedding $e$. The whole process is applied for both the documents $d_1$ and $d_2$ to get the aggregated embeddings $e_1$ and $e_2$ from their textual embeddings $t_1$, $t_2$ and network embeddings $n_1$, $n_2$ respectively. The final similarity between $d_1$ and $d_2$ is the cosine similarity between embeddings $e_1$ and $e_2$. 
Both the textual and network embeddings are of the same dimension ($dim = 200$) and are $L_2$ normalized. 
We experimented with $dim=100$ and $200$, and observed better performance with $dim=200$ for both textual and network-based methods, on the validation set.

We now describe various approaches we take to combine the text and network embeddings of a single document, specifically the different methods for computing $f(t,n)$.

% Note that this approach is a slight departure from the general scheme shown in Figure~\ref{fig:workflow} where we first compute the textual similarity (using $t_1$, $t_2$) and network similarity (using $n_1$, $n_2$) and then combine the two scores. 

\vspace{3mm}
\noindent \textbf{(II) Unsupervised Embedding Combinations}: Here we explore simple unsupervised techniques for combining the textual embedding ($t_1$ for $d_1$ and $t_2$ for $d_2$) and the network embedding ($n_1$ for $d_1$ and $n_2$ for $d_2$) to produce a resultant embedding of a document. From our running example, let the resultant embedding of $d_1$ and $d_2$ be $e_1$ and $e_2$ respectively. The similarity between the document pair $(d_1,d_2)$ is $cosine\_similarity(e_1,e_2)$. Now, we describe the methods for obtaining $e_1$ and $e_2$. \\
\underline{(i) Emb-Average} -- here the final embedding of a document $d$ is the element-wise average of $t$ and $n$. Therefore, $e_1\!=\! t_1 \oplus n_1$  and $e_2\!=\! t_2 \oplus n_2$, where $\oplus$ denotes element-wise averaging. Note that, the dimension of $e_1$ and $e_2$ is $dim$. \\
\underline{(ii) Emb-Max} -- here the final embedding of a document is the element-wise maximum of $t$ and $n$. Thus $e_1\!=\! max(t_1, n_1$)  and $e_2\!=max(t_2, n_2)$, where $max$ denotes element-wise maximum. The dimension of $e_1$ and $e_2$ is $dim$. \\
\underline{(iii) Emb-Conc~\cite{conc}} -- here, the final embedding of a document is the concatenation of $t$ and $n$. Therefore, $e_1\!=\! t_1 \odot  n_1$  and $e_2\!=t_2 \odot n_2$, where $\odot $ denotes concatenation operation. Note that, the dimension of $e_1$ and $e_2$ is $2 \times dim$.

\vspace{3mm}
\noindent
\textbf{(III) Neural Network based Self-supervised Combinations:} Neural network based methods have been widely used for learning multi-modal embeddings for an entity from text and visual/audio information. 
For instance, Talleda et al.~\cite{nn} used text and visual modalities while Wang et al.~\cite{autoencoder} used text, audio and visual modalities. Both these works aimed to improve word representations that were evaluated on word similarity and relatedness tasks. 
In the present work, we adapt the two supervised techniques -- NN-mapping~\cite{nn} and Auto-encoder~\cite{autoencoder} -- for learning multi-modal embeddings for a document $d$ from its text and network information ($t$ and $n$ respectively). 

\vspace{2mm}
In the neural network based mapping (NN-mapping) approaches, the objective is to project representations from the distributed space of one modality (e.g., text) into the distributed space of another modality (e.g., network). In this architecture, the input is the actual textual embedding $t$ and the actual network embedding $n$ of a document $d$. The output is a `imagined/predicted' embedding $n'$ in the network modality. $t$ is passed through fully connected layers that learns a mapping function $M$ from text modality to the network modality. The loss function is calculated as Mean Squared Error (MSE) Loss between the `imagined/predicted' network embedding $n' = M(t)$ and the actual network embedding $n$ of the same document, i.e.,  $loss (n,n') = \frac{1}{2} {\left \|n' - n \right \|}{_2}{^2}$. 
During inference, for document $d_1$ and $d_2$, we get the `imagined/predicted' network embeddings $n'_1 \! = \! M(t_1)$ and $n'_2 \! = \! M(t_2)$. Then, the mapped embeddings $n'_1$ and $n'_2$ are concatenated with the original textual embeddings $t_1$ and $t_2$ respectively for $d_1$ and $d_2$ in two ways -- \\
(i) \underline{NN-map+Conc}: In this approach~\cite{nn}, the mapped embeddings and original textual embeddings are concatenated, because the mapped embeddings are essentially `imagined/predicted network embeddings' and hence textual embeddings should be concatenated to have the complete multimodal representation. The resultant embedding of $d_1$ is $e_1 \! = \! n'_1 \odot t_1 $ and that of $d_2$ is $e_2 \! = \! n'_2 \odot t_2 $. Finally, $cosine\_similarity(e_1,e_2)$ is computed for getting the similarity of $(d_1, d_2)$; 

\vspace{2mm}
\noindent (ii) \underline{NN-Map+Wtd.Conc}: In this approach~\cite{conc}, instead of simply concatenating the embeddings, an optional tuning hyper-parameter $\alpha$ is used as proposed in~\cite{conc}. Hence, the resultant embedding of $d_1$ is $e_1 \! = \! \alpha \! \times \! n'_1 \odot (1-\alpha)\! \times t_1$ and that of $d_2$ is $e_2 \! = \! \alpha \! \times \! n'_2 \odot (1-\alpha)\! \times t_2$. The similarity between $(d_1,d_2)$ is computed as $cosine\_similarity(e_1,e_2)$.

\vspace{2mm}
\noindent \underline{(iii) AutoEncoder:}
In this approach~\cite{autoencoder}, the text embedding $t$ and network embedding $n$ of a document is provided in both the input and the output. The text embedding and network embedding are separately encoded through stacked, denoising auto-encoders. These encoded vectors are then concatenated in the multi-modal layer, $L$. From this concatenated layer, individual text and network embeddings are reconstructed/decoded. The loss here is also Mean Squared Error, computed as $loss (t,n,t',n') = {\left \|t' - t \right \|}{^2} +  {\left \|n' - n \right \|}{^2}$, where $t'$ and $n'$ are the reconstructed text and network embeddings respectively. During the inference phase, both the text embedding $t$ and network embedding $n$ of a document $d$ are passed through the encoder layers and the multi-modal layer $L$ to obtain the multi-modal embedding $e$. Hence, the multi-modal embedding for $d_1$ is $e_1 \! = \! L(t_1,n_1)$, and that for $d_2$ is $e_2\! = \!L(t_2, n_2)$. Similarity between $(d_1,d_2)$ is computed as $cosine\_similarity(e_1,e_2)$.

\vspace{2mm}
\noindent \underline{Training the neural combination models:}
A big advantage of the neural combination models described above is that they are self-supervised, and can be trained using the text embedding and network embedding of the same document (as described above). 
Among the $30,056$ documents for which we have both textual and network embeddings (see Section~\ref{ss:hierspcnet}), we consider all documents excluding those in the validation and test sets. This set of documents is split into 80\%:20\% train:validation (this validation set is specifically used for determining the optimal hyper-parameter values for the neural combination models).  

\vspace{1mm}
\noindent \underline{Implementation details:} We implement the NN-map models following the description in~\cite{nn}, using a two-layer fully connected neural network with 250 and 300 neurons.
For the NN-Map+Wtd.Conc method, we consider $\alpha$ as 0.5.
For the Autoencoder approach (implementation publicly available at \url{https://github.com/wangshaonan/Associative-multichannel-autoencoder}), the textual and network encoders are of sizes $150$ and $100$, and the size of the multi-modal layer is $300$. We use the AdamW optimizer and learning rate of $0.01$ for all the neural combination methods.
All these hyper-parameter values are decided based on the validation set for the neural combination models; 
we consider that model which gives the least MSE loss on this validation set.
%We use that model over the test set, that gives the best performance over the validation set.

\subsection{Graph-based methods for combining network and text signals}

Here we use graph-based approaches for combining the text and network representations of a given document-pair, for computing the similarity between the documents. We try the following two approaches:

\vspace{2mm}
\noindent \underline{(i) Using Hier-SPCnet}: In this approach, we explore state-of-the-art Node Representation Learning methods that take as input a graph $G=(V,E,X)$ where V is the vertex set, E is the edge set and X is a matrix of node features in $\mathbb{R}^{m \times \left | V \right |}$ where every vertex $v \in V$ is represented with an $m$-dimensional feature vector. Given such a graph, node representation learning algorithms aim to learn embedding of the vertices in $\mathbb{R}^{d}$, where $d$ is the node embedding dimension.
In our setting, the graph $G$ is Hier-SPCNet described in Section~\ref{ss:hierspcnet}. 
In Hier-SPCnet, nodes are  documents (for which we intend to compute similarity), sections, parts, chapters, topics and acts. The node feature matrix $X$ consists of the textual embedding $t$ of the documents. 

It can be noted that the Statute nodes, specifically the Section/Article nodes, in Hier-SPCNet also have textual content (see Figure~\ref{fig:workflow}). We learn another Doc2Vec model on the statutes, through which we infer the textual embeddings of section, part, chapter, topic and act nodes. Note that, in a Statute (e.g., Constitution of India, 1950), textual matter is present only in the section (leaf) nodes (e.g., Article 21 of the Constitution of India, 1950). To provide a feature vector for the internal nodes (topic, chapter, part) and the root node (act) as well, we recursively add the textual embedding of its child nodes. 

Given this setting, we use three state-of-the-art methods for Node Representation Learning --  TADW~\cite{tadw}, GCN~\cite{gcn} and GraphSAGE~\cite{graphsage} -- that takes the above graph as input and outputs node embeddings $e$ that takes into account both the textual content $t$ given as the node feature, as well as the network structure. For computing similarity between the document pair $(d_1,d_2)$, we compute $cosine\_similarity(e_1,e_2)$ where $e_1$ and $e_2$ are the learnt representations of the document nodes $d_1$ and $d_2$ respectively.

\vspace{2mm}
\noindent \underline{(ii) Using a network constructed from text} We also explore Paper2Vec~\cite{paper2vec} that constructs a graph with nodes as documents and edges exist between two documents if their textual similarity is above a particular threshold. 
DeepWalk is then applied on this network to get the node (here, document) embeddings.

\vspace{2mm}
\noindent \underline{Implementation details:} We implement Paper2Vec following the description in~\cite{paper2vec} as closely as possible, with text similarity threshold $0.5$, node embedding dimension $200$ and other parameters set to what is stated in~\cite{paper2vec}. 
We use the publicly available implementations of TADW \footnote{\url{https://github.com/thunlp/OpenNE/}}, GCN \footnote{\url{https://stellargraph.readthedocs.io/en/stable/demos/node-classification/gcn-node-classification.html}} and GraphSAGE \footnote{\url{https://stellargraph.readthedocs.io/en/stable/demos/embeddings/graphsage-unsupervised-sampler-embeddings.html}}. 
TADW was run for 30 iterations. 
GCN and GraphSAGE used 2 hidden layers with $128$ and $64$ hidden units respectively, with relu activation and was run for 50 iterations. We use all the 4 aggregation functions of GraphSAGE; we report results for Mean which gave the best result.

%% file: sections/results-text+network.tex
We apply all Text + Network Combination methods described above on the validation and test sets of 100 and 90  document-pairs respectively for which we have gold standard similarity scores from experts. Then we measure correlation, MSE, and F-score for each method. The results over the validation set are reported in Table~\ref{tab:main-results} and the results over the test set are in Table~\ref{tab:main-results-test}. We repeat the performances of the best text-based method (Doc2Vec) and the best network-based method (Hier-SPCNet-ICF-m2v) in Tables~\ref{tab:main-results} and~\ref{tab:main-results-test}  for ease of reference.

We observe that the performances of most methods are slightly lower over the test set than over the validation set; this is somewhat expected, since the hyper-parameters of the methods are optimized over the validation set and then the same hyper-parameter values are directly applied over the test set (no further tuning is carried out over the test set).

\begin{table}[t]
\centering
\caption{Results of the different methods for computing legal case document similarity, on the \textbf{validation set}. The overall best value for each metric is in \textbf{bold}; best value within the group of methods using a particular combination function is \underline{underlined}. The first two rows are repeated from earlier sections, for ease of reference.}
\label{tab:main-results}
%\resizebox{\columnwidth}{!}{
\begin{tabular}{|c|c|c|c|c|c|}
\hline
\textbf{Type}  & \textbf{Combination Function}  & \textbf{Method}   & \textbf{Correlation} & \textbf{MSE}    & \textbf{Fscore}  \\ \hline
Text based & --  &  Doc2Vec & 0.765 & 0.039 & 0.768 \\ \hline \hline
Network based & -- & Hier-SPCNet-ICF-m2v & 0.725 & 0.0427 & 0.779 \\ \hline \hline
\multirow{13}{*}{\textbf{\begin{tabular}[c]{@{}c@{}}Text\\ +\\ Network\\ \\ Combination\end{tabular}}} & \multirow{2}{*}{\begin{tabular}[c]{@{}c@{}}Value\\ combination\end{tabular}}      & Value-Average        & \underline{0.774}          & 0.0401    & \textbf{0.825}       \\ \cline{3-6} 
  & & Value-Max & 0.756 & \underline{0.0381} & 0.778 \\ \cline{2-6} \cline{2-6} 
  & \multirow{3}{*}{\begin{tabular}[c]{@{}c@{}}Unsupervised Embedding\\ Combination\end{tabular}}  & Emb-Average & 0.742 & \underline{0.0389}    & 0.801  \\ \cline{3-6} 
  & & Emb-Max  & 0.741  & 0.0624 & 0.700  \\ \cline{3-6} 
  & & Emb-Conc & \underline{0.775} & 0.0399 & \textbf{0.825} \\ \cline{2-6} \cline{2-6}
  & \multirow{4}{*}{\begin{tabular}[c]{@{}c@{}}Graph-based\\ combination\end{tabular}}  & Paper2Vec & 0.642 & \underline{0.0614} & \underline{0.725}  \\ \cline{3-6} 
  &     & TADW  & \underline{0.696} & 0.0625 & 0.686    \\ \cline{3-6} 
  &     & GCN   & 0.331 & 0.2472 & 0.394   \\ \cline{3-6} 
  &     & GraphSage     & 0.583 & 0.1827 & 0.652   \\ \cline{2-6} \cline{2-6} 
 % & \multirow{4}{*}{\begin{tabular}[c]{@{}c@{}}Neural\\ Network\\ based\end{tabular}} & NN-Map  & 0.796 & 0.0607 & 0.767\\ \cline{3-6} 
  &  \multirow{3}{*}{\begin{tabular}[c]{@{}c@{}}Neural Network\\ based combination\end{tabular}}    & NN-Map+Conc  & \textbf{0.804} & \textbf{0.0349} & \underline{0.820}   \\ \cline{3-6} 
  &     & NN-Map+Wtd.Conc    & 0.761 & 0.0493  & 0.682 \\ \cline{3-6} 
  &     & AutoEncoder        & 0.704 & 0.2563  & 0.315  \\ \hline
\end{tabular}
\end{table}

\begin{table}[t]
\centering
\caption{Results of the different methods for computing legal case document similarity, on the \textbf{test set}. The overall best value for each metric is in \textbf{bold}; best value within the group of methods using a particular combination function is \underline{underlined}. The first two rows are repeated from earlier sections, for ease of reference.}
\label{tab:main-results-test}
\begin{tabular}{|c|c|c|c|c|c|}
\hline
\textbf{Type}  & \textbf{Combination Function}  & \textbf{Method}   & \textbf{Correlation} & \textbf{MSE}    & \textbf{Fscore}  \\ \hline
Text based & --  &  Doc2Vec & 0.701 & 0.0356 & 0.682 \\ \hline \hline
Network based & -- & Hier-SPCNet-ICF-m2v & 0.650 & 0.0402 & 0.665 \\ \hline \hline
\multirow{13}{*}{\textbf{\begin{tabular}[c]{@{}c@{}}Text\\ +\\ Network\\ \\ Combination\end{tabular}}} & \multirow{2}{*}{\begin{tabular}[c]{@{}c@{}}Value\\ combination\end{tabular}}      & Value-Average        & \underline{0.716}          & \underline{0.0378}    & \underline{0.731}       \\ \cline{3-6} 
  & & Value-Max & 0.704 & 0.0384 & 0.701 \\ \cline{2-6} 
  & \multirow{3}{*}{\begin{tabular}[c]{@{}c@{}}Unsupervised Embedding\\ Combination\end{tabular}}  & Emb-Average & 0.674 & \underline{0.0349}    & 0.711  \\ \cline{3-6} 
  & & Emb-Max  & 0.739  & 0.0471 & 0.664  \\ \cline{3-6} 
  & & Emb-Conc & \underline{0.743} & 0.0362 & \textbf{0.743} \\ \cline{2-6}
  & \multirow{4}{*}{\begin{tabular}[c]{@{}c@{}}Graph-based\\ combination\end{tabular}}  & Paper2Vec & 0.582 & 0.0672 & \underline{0.623}  \\ \cline{3-6} 
  &     & TADW  & \underline{0.621} & \underline{0.0615} & 0.614    \\ \cline{3-6} 
  &     & GCN   & 0.304 & 0.2972 & 0.331   \\ \cline{3-6} 
  &     & GraphSage     & 0.507 & 0.2227 & 0.596   \\ \cline{2-6} \cline{2-6}
 % & \multirow{4}{*}{\begin{tabular}[c]{@{}c@{}}Neural\\ Network\\ based\end{tabular}} & NN-Map  & 0.796 & 0.0607 & 0.767\\ \cline{3-6} 
  &  \multirow{3}{*}{\begin{tabular}[c]{@{}c@{}}Neural Network\\ based combination\end{tabular}}    & NN-Map+Conc  & \textbf{0.784} & \textbf{0.0326} & \underline{0.740}   \\ \cline{3-6} 
  &     & NN-Map+Wtd.Conc    & 0.752 & 0.0403  & 0.708 \\ \cline{3-6} 
  &     & AutoEncoder        & 0.725 & 0.2063  & 0.395  \\ \hline
\end{tabular}
\vspace{-5mm}
\end{table}

The simple \textit{value combination methods}, especially Value-Average, perform quite well over both validation and test sets. In fact, Value-Average achieves the joint-highest F-score (0.825) across all methods over the validation set, and also performs quite well over the test set (F-score 0.731).

The unsupervised embedding combination method Emb-Conc (which is a concatenation of the normalized text and network embeddings) performs the best in terms of F-score among all methods over both the validation set (joint highest F-score 0.825) as well as over the test set (F-score 0.743). However, other methods out-perform Emb-Conc in terms of the metrics Correlation and MSE.

We see that the Graph-based combination methods (out of which TADW performs the best) do not perform well -- their performance is worse than the best individual text-based (Doc2Vec) and network-based (Hier-SPCNet-ICF-m2v) methods. This is because the graph-based methods considers Hier-SPCNet to be just another network. The random walks are not \textit{guided} in the way it is done in metapath2vec through metapaths. 
%This approach is promising and may open up future research directions to combine textual and network embeddings. 

We find the neural network-based combination method NN-Map+Conc to give the best performance across all methods over both the validation and test sets, in terms of both Correlation (0.804 over validation set and 0.784 over test set) and MSE. 
NN-Map+Conc shows 11.8\% and 20.6\% improvement in terms of correlation with expert scores, over the best text-based method (Doc2vec) and the best network-based method (Hier-SPCNet-ICF-m2v) respectively over the test set. These improvements are statistically significant by paired Student's T-Test at 95\%, $p<0.05$.
In terms of F-score, NN-Map+Conc achieves the second-highest F-Score (0.740) over the test set  which is very close to the highest F-score of 0.743 (achieved by Emb-Conc).

Thus, while NN-Map+Conc performs the best in terms of correlation and MSE, some of the simpler combination methods (notably, Emb-Conc) are seen to perform slightly better in terms of the F-score metric. 
The F-score metric considers a 2-class classification of document-pairs as similar or dissimilar. 
If such a classification is sufficient for a particular application, then Emb-Conc would probably be the preferred method due to its relative simplicity.
However, just a binary classification of document-pairs may {\it not} be sufficient for certain applications, e.g., for a retrieval/recommender system which wants to output a \textit{ranked list} of similar documents for a given query/source document. 
Such an application will require a numerical similarity measure for a document-pair, based on which retrieved documents can be ranked.
For such a ranking application, NN-Map+Conc is more useful since it achieves the best correlation and MSE with respect to expert-assigned similarity scores.

%% file: sections/reco-study.tex
% \begin{table}[t]
% \caption{Quality of top-3 recommendations generated by Emb-Conc and NN-Map+Conc, as judged by legal experts. The Algorithmic Sim and Expert Sim values are averaged over all 45 recommendations by a particular method. Almost all the recommended case documents are judged to be citable from the source case documents.}
% \label{tab:reco-results}
% \begin{tabular}{|c|c|c|}
% \hline
% \textbf{Evaluation Metrics} & \textbf{NN-Map+Conc} & \textbf{Emb-Conc} \\ \hline
% \textbf{Algorithmic Sim} & 0.747 & 0.704 \\ \hline
% \textbf{Expert Sim} & 0.757 & 0.714 \\ \hline
% \textbf{Standard Deviation} & 0.0808 & 0.1502 \\ \hline
% \textbf{MSE} & 0.0128 & 0.0472 \\ \hline \hline 
% \textbf{\% citable by $1^{st}$ Law Expert} & 86.66 (39/45) & 75.55 (34/45) \\\hline
% \textbf{\% citable by $2^{nd}$ Law Expert} & 97.77 (44/45) & 86.66 (39/45) \\\hline
% \end{tabular}
% \end{table}

From our discussion with Law experts, we understand that, in a case document, \textit{not all} relevant/similar cases are cited.
This is primarily because of the limited time of the Law practitioners who write the case documents. 
Generally, only the prior-cases that directly influence/form a part of the rationale of the court for the final verdict of the current case, are cited. 
%This is in unlike how papers are cited in related work -- authors tend to cite \textit{almost all} related papers that are similar in some way to the current work. 
We find that on average, a case document cites 4 precedent cases. But there are many other similar case documents that are left uncited.
While studying a particular case document (say, $s$), retrieving/identifying case documents  that are {\it not actually cited} from $s$, but are so similar to $s$ that they {\it could have been cited} from $s$, is a practically useful problem. Identifying such `missing' citable case is very important for a Law practitioner / academician, e.g., for challenging the verdict of the case $s$ in a higher court (for a law practitioner) and gaining knowledge about a topic of law (for a law academician). 
While most popular commercial legal information systems show the cases that are actually cited from a source case document $s$, to our knowledge, there is no system presently that identifies cases that are citable (sufficiently similar) from $s$ but are not actually cited.

In the previous section, we saw that the similarity values predicted by the NN-Map+Conc method match the best with expert-assigned similarity scores (leading this method to achieve the best correlation and MSE values). 
We now explore the utility of the method for this practical application of retrieving / recommending uncited but similar case documents for a given source / query document.

\vspace{3mm}
\noindent \textbf{Experiment:} We choose a source case document (say, $s$) and use the NN-Map+Conc method to output a ranked list of all Indian Supreme Court cases that are {\it not cited already from $s$}, in decreasing order of their similarity with $s$. From this list, we select the top 3 documents (Top-3) that are retrieved/recommended to be most similar to $s$ by NN-Map+Conc.
%2 documents from the middle of the list (Mid-2) and two documents from the end of the list (Last-2).
Two law experts are shown the source case $s$ and each of the Top-3 retrieved case documents, and asked to rate the quality of each retrieved result through two measures -- 
(i)~\textit{sim}: Each expert is asked to assign a similarity score in $[0,1]$ indicating how similar $s$ is with a recommended case ($1$ indicates maximum similarity), and 
(ii)~\textit{citable}:  We ask the law experts -- could the recommended case have been cited from the source case $s$ (or vice versa, depending on the chronological ordering)? The law experts give a 0/1 answer to this question, where 1 implies that the recommended case could have been cited, 0 otherwise. 

We randomly select $15$ source documents, and for each source document, we evaluate the Top-3 documents (that are not already cited from the source document) retrieved/recommended by the method NN-Map+Conc. Out of these $15$ source documents, $9$ documents were from the same validation/test set that was used in the earlier sections. The remaining $6$ documents were sampled from the remaining pool of Indian Supreme Court case documents. 
Thus, we get $15 \times 3 = 45$ Top-3 recommendations of the NN-Map+Conc algorithm evaluated by two law experts.

\begin{table}[t]
\centering
\caption{Quality of Top-3 recommendations generated by  NN-Map+Conc, as judged by two law experts. The {\it Algorithmic Sim} (estimated by the method) and {\it Expert Sim} (assigned by experts) values are averaged over all recommendations that were evaluated. Almost all the Top-3 recommended case documents are judged to be citable from the source case documents.}
\label{tab:reco-results}
\begin{tabular}{|c|c|}
\hline
\textbf{Evaluation Metrics} & \textbf{Top-3} \\ \hline
\textbf{Average Algorithmic Sim} & 0.747  \\ \hline
\textbf{Average Expert Sim} & 0.757  \\ \hline
\textbf{Standard Deviation} & 0.0808 \\ \hline
\textbf{MSE} & 0.0128 \\ \hline \hline 
\textbf{\% judged citable by $1^{st}$ Law Expert} & 86.66 (39/45) \\\hline
\textbf{\% judged citable by $2^{nd}$ Law Expert} & 97.77 (44/45) \\\hline
\end{tabular}
\vspace{-4mm}
\end{table}

\vspace{3mm}
\noindent \textbf{Evaluation:} 
We use the term \textit{Algorithmic Sim} for the similarity values inferred by NN-Map+Conc. The term \textit{Expert Sim} is used for the average similarity value assigned by the two legal experts, 
Table~\ref{tab:reco-results} shows that the top 3 recommendations by NN-Map+Conc receive an average similarity score of $0.757$ from the experts, while their algorithmic similarity was $0.747$ on average. To understand how different the scores given by the annotators and estimated by NN-Map+Conc vary, we also compute Standard Deviation and Mean Square Error (MSE).\footnote{Unlike in the previous evaluation where there were  document-pairs with widely varying similarity (spread out almost uniformly in [0,1]), the present evaluation is being carried out over document-pairs that are actually very similar, e.g., all document pairs considered in this evaluation have algorithmic similarity higher than $0.7$. This is why there is no linearity among the similarity scores in this evaluation; hence we do not report correlation.} 
We observe that both the values are fairly low, suggesting the utility of the method in generating good quality recommendations. 
Importantly, as shown in the last two rows of Table~\ref{tab:reco-results}, \textbf{a large fraction of the Top-3 recommendations by NN-Map+Conc are judged citable by both law experts (39/45 and 44/45 respectively)}.

Table~\ref{tab:reco-egs} shows three example source documents and their Top 3 recommendations by the method. We find that the difference in the expert scores is not more than $0.2$, suggesting that the experts agree in general. Also both the experts judge the top recommendations given by NN-Map+Conc as very similar to the source documents.

Therefore, we conclude that the method developed in this work has an important practical utility to the law practitioners for retrieving/recommending citable (similar but uncited) cases for a given source case document. 

% We do not report Pearson Correlation with Expert score and the algorithmically generated scores because there is no linearity among these scores. Our algorithm recommend top 3 documents that have the highest similarity with the source document. The experts in general find the recommended documents to be similar but may find the third recommended document to be most similar than the first two, for instance. Therefore correlation is not a suitable metric for evaluation in this setting. 

\vspace{1mm}
\noindent 

\begin{table}[tb]
\caption{Top 3 Recommendations given by NN-Map+Conc for three source documents, along with the similarity scores given by the two law experts. The experts judge the top recommendations to be very similar to the source documents.}
\label{tab:reco-egs}
\begin{tabular}{|c|p{0.5\columnwidth}|c|c|}
\hline 
%\multirow{2}{*}{\textbf{Source Doc}} & \multicolumn{3}{c|}{Recommendations by NN-Map+Conc} \\ \cline{2-4} 
{\bf Source Doc} & {\bf Recommendations by NN-Map+Conc} & \textbf{Expert 1} & \textbf{Expert 2} 
\\ \hline \hline

\multirow{3}{*}{\begin{tabular}[c]{@{}c@{}}Raju @Devendra Choubey \\ v \\ State of Chhatisgarh\end{tabular}} & Venkatesan v State of Tamil Nadu & 0.7 & 0.8 
  \\ \cline{2-4} 
  & State of Madhya Pradesh v Chamru @ Bhagwandas & 0.8 & 1.0 
  \\ \cline{2-4} 
 & Md. Kalam @ Abdul Kalam v State of Rajasthan & 0.7 & 0.9 
\\ \hline \hline

\multirow{3}{*}{\begin{tabular}[c]{@{}c@{}}Inspector of Police, Tamil Nadu \\ v Balaprasanna\end{tabular}} & Pannayar v State of Tamil Nadu, By Inspector of Police (2009) & 0.6 & 0.8 
 \\ \cline{2-4} 
 & Raja @ Rajinder v State of Haryana & 0.7 & 0.9 
 \\ \cline{2-4} 
 & Sanatan Naskar \& Anr. v State Of  West Bengal & 0.7 & 0.9 \\ \hline \hline
 
\multirow{3}{*}{\begin{tabular}[c]{@{}c@{}}Kelvinator of India Limited \\ v \\ State of Haryana\end{tabular}} & M/s Sahney Steel and Pressworks Ltd v The Commercial Tax Officer And Others & 0.7 & 0.7 \\ \cline{2-4} 
 & M/S Hyderabad Engineering Industries v State Of Andhra Pradesh & 0.8 & 0.9 \\ \cline{2-4} 
 & Union of India and Another v K. G. Khosla and Company Ltd and Others & 0.7 & 0.8 \\ \hline
\end{tabular}
\vspace{-5mm}
\end{table}

%% file: sections/conclusion.tex
In this paper, we substantially improved the state-of-the-art for estimating similarity between legal case documents. We first propose Hier-SPCNet and incorporate domain knowledge on this network for efficient network-based similarity (Hier-SPCNet-ICF-m2v). 
We attempt to combine the text and network similarity signals intelligently, which to the best of our knowledge has not been tried earlier, especially for legal document similarity. 
We find NN-Map+Conc to be performing quantitatively the best in terms of correlation between the predicted similarity values and expert-assigned similarity values. We investigate the practical utility of the method to recommend case documents to law experts. The experts appreciated the recommendations and found them to be useful in recommending citable prior cases. 
%Thus we establish NN-Map+Conc to be the best method for legal document similarity that performs both quantitatively and qualitatively best. 

It can be noted that different models considered in this work take different times to train. Hier-SPCNet-ICF-m2v took 10-12~hours to train in our setting. Doc2Vec got trained in only 1~hour. NN-Map+Conc took only about 1~hour to train once the text and network embeddings are available. While it may be difficult to run the method proposed in this work completely in an online setting (primarily due to the large time taken by Hier-SPCNet-ICF-m2v), it can be easily used to generate real-time recommendations as follows.
The method can be executed offline, and the similarity values of all document-pairs can be stored in a database. When a query (a source document) comes, the system can directly use the pre-computed similar values and display the top $k$ most similar documents as recommendations in real-time.

\vspace{2mm}
\noindent {\bf Future directions of work:}
There are several opportunities to extend the work reported in this paper. 
For instance, in this work, we do not explore how recommendations can be provided for an {\it unseen} document, which is not present as a node in Hier-SPCNet, in a dynamic setting; this is left as a future work.
Also, we plan to improve the text-based similarity estimation methods by using domain-specific knowledge. In particular, we observed that Doc2vec trained over Indian legal documents enables better similarity estimation over Indian legal documents, than Bert-based models pre-trained over other types of legal data (e.g., LegalBert pre-trained over European and US legal documents). 
Hence, in future, we plan to pre-train transformer models on Indian legal documents and check if the estimation of similarity can be improved.
Additionally, we plan to explore better ways of developing large-scale training datasets for \textit{supervised} similarity estimation models, and check if such supervised models (suitably trained) can achieve better similarity estimation.

Also we would like to explore other applications of our similarity methods, e.g., building a \emph{Legal Semantic Network} (similar to the \emph{Semantic Web}), densification of the legal citation networks that are known to be very sparse, clustering legal documents, and so on. 
We will also attempt to add explanations to the estimated similarity values, which is strongly desired by the law practitioners.

Finally, we have experimented only on Indian legal documents in this paper, and an important future work is to make the algorithms generalizable to legal case documents from different countries.
The algorithms developed in this paper assume that the legal case documents contain text, citations to relevant prior-cases (i.e., other legal case documents) and citations to relevant statutes (or laws of a particular jurisdiction). Legal court case documents in many countries and jurisdictions generally have these properties. As such, we believe that the algorithms developed in this paper can be applied to documents of any Common Law jurisdiction
that can be modelled as a citation network (e.g., that of France~\cite{Mazzega:2009:NFL:1568234.1568271}).
The text-based methods used in this paper can be used for any other country, if models such as Doc2vec or transformer models are trained over legal documents of that country. 
For the network-based methods to be applied over legal documents of another country, 
the Hier-SPCNet needs to be developed for that country. For this, several steps would need to be taken -- the statute (laws/legislations) citations and prior-case citations need to be extracted from legal documents of that country. Alongside, the Metapaths that encode domain knowledge about document similarity have to be developed in consultation with law experts from that country. 
We plan to follow these steps meticulously and apply our algorithms to legal documents of other Common Law countries in future.